\documentclass[a4paper,12pt]{article} 
\usepackage{graphicx}
\usepackage{amssymb}
\usepackage{amsmath}
\usepackage[colorlinks=true, pdfstartview=FitV, linkcolor=blue,  citecolor=blue, urlcolor=blue]{hyperref}  
\usepackage{float}
\graphicspath{ {images/} }

\newcommand{\dfr}{\widehat{d}} 

\newcommand{\Vsc}{{\widehat V}} 
\newcommand{\Ds}{{\widetilde D}} 

\def\l{\left}
\def\r{\right}
\def\beq{\begin{equation}}
\def\eeq{\end{equation}}
\def\d{\partial}

\def\beq{\begin{equation}}\def\eeq{\end{equation}}
\def\bea{\begin{eqnarray}}\def\eea{\end{eqnarray}}

\usepackage{parskip}

\begin{document}

\title{The origins of macroscopic quantum coherence in high temperature superconductivity}

\author{Philip Turner$^1$\footnote{ph.turner@napier.ac.uk}  and Laurent Nottale$^2$\footnote{laurent.nottale@obspm.fr}\\
$^1${\small Edinburgh Napier University, 10 Colinton Road, Edinburgh, EH10 5DT, United Kingdom.} \\
$^2${\small CNRS, LUTH, Observatoire de Paris-Meudon, 5 Place Janssen, 92190, Meudon, France.}} 
\date{April 27th, 2015}
\maketitle

\begin{abstract}
A new, theoretical approach to macroscopic quantum coherence and superconductivity in the $p$-type (hole doped) cuprates is proposed.  The theory includes mechanisms to account for e-pair coupling in the superconducting and pseudogap phases and their inter relations observed in these materials.  

Electron pair coupling in the superconducting phase is facilitated by local quantum potentials created by static dopants in a mechanism which explains experimentally observed optimal doping levels and the associated peak in critical temperature.  By contrast, evidence suggests that electrons contributing to the pseudogap are predominantly coupled by fractal spin waves (fractons) induced by the fractal arrangement of dopants.    

On another level, the theory offers new insights into the emergence of a macroscopic quantum potential generated by a fractal distribution of dopants.  This, in turn, leads to the emergence of coherent, macroscopic spin waves and a second associated macroscopic quantum potential, possibly supported by charge order.  These quantum potentials play two key roles.  The first involves the transition of an expected diffusive process (normally associated with Anderson localization) in fractal networks, into e-pair coherence.  The second involves the facilitation of tunnelling between localized e-pairs.  These combined effects lead to the merger of the super conducting and pseudo gap phases into a single coherent condensate at optimal doping.  The underlying theory relating to the diffusion to quantum transition is supported by Coherent Random Lasing, which can be explained using an analogous approach.   As a final step, an experimental program is outlined to validate the theory and suggests a new approach to increase the stability of electron pair condensates at higher temperatures.  
\end{abstract}
\textbf{Keywords:}
High temperature superconductivity, $p$-type cuprates, macroscopic quantum coherence, electron pairing, fractal networks. 

\subsection*{\centering{1. Introduction}}  
In conventional superconductors (SC), electrons are bound through interactions between electrons and quanta of lattice vibrations (phonons), to form a bosonic fluid of electron-pairs (e-pairs), in a mechanism first described by Bardeen, Cooper, and Schrieffer (BCS) \cite{Bardeen1957A, Bardeen1957B,DeGennes1989}.  The relatively long range coherence length of e-pairs is influenced by material characteristics but also by an increase in the de Broglie wave length $\lambda_{deB}=\hbar/p$ as momentum decreases with decreasing temperature.  The coherence length permits large numbers of e-pairs to occupy the same density of states (DOS) creating a lower entropy, phase coherent, macroscopic quantum state with long range order.  However, the energy gap $E_g$ between condensed and normal states (due to low phonon coupling energies) is small, resulting in low critical temperatures $T_c$ ($\lesssim20$ K) when compared with High Temperature Super Conducting (HTSC) materials.  After more than two decades of intense experimental and theoretical research, there is still no widely accepted theory to explain macroscopic quantum coherence and the significantly higher electron coupling energies in these high temperature superconductors.  In this paper we review a number of studies offering some key insights, which we consider in the development of a new theoretical approach to explain HTSC properties.  

Due to the enourmous number of papers published on a range of different materials with different characteristics, we have focussed on the most studied $p$-type (hole doped) family of cuprates.  However, where appropriate we have indicated how we think the theory may have applicability to the broader family of HTSC materials.

\subsection*{\centering{2. Background}}
\subsection*{\centering{2.1. Structural geometry and its influences}}

The symmetry of a structure on its microscopic scale is an important property of condensed matter systems.  Fluids, glasses or amorphous solids have dispersion relations with a rotational symmetry, whilst crystalline materials are anisotropic, leading to a spatial anisotropy of DOS, with particle momentum being geometrically constrained. One of the more notable aspects of HTSC materials lies in their disordered structure with a number of papers \cite{Apalkov2002,Buttner1987,Fratini2010,Hufner2008,McElroy2005,Milovanov2002,Tranquada2004, Poccia2011,Poccia2012}, indicating that HTSC is favoured by complex fractal systems. A key objective of this current work is to explore a possible link between the geometry of these materials and their unusual properties.

As part of the discussion of quantum coherence in disordered materials we include the recently discovered phenomena of Coherent Random Lasing (CRL), which occurs in fractal media under very specific geometries \cite{Apalkov2002}.  As with HTSC, no single theory has yet been universally accepted to explain CRL.  However, there are some obvious parallels between the two phenomena, with a clear link between network geometry and macroscopic quantum coherence in bosonic fluids.

Considering a Schr\"odinger equation describing electrons moving in a random potential, and the scalar wave equation for light propagation in a medium with fluctuating dielectric constant \cite{Apalkov2002}, both equations can be presented in a generic form
\beq
\Delta\psi(x)+[k-U(x)]\psi(x)=0   															\label{eq.1}
\eeq
For an electron with mass $m$ and energy $E$, moving in a random potential $V(x)$, the parameters $k$ and $U(x)$ are
$k=2mE$, $U(x)=2mV(x)$.
For a light wave with frequency $\omega$, traveling in a medium with dielectric constant $\epsilon$, corresponding expressions for $k$ and $U(x)$ take the form
$k=\epsilon(\frac{\omega}{c}),    U(x)=-\delta\epsilon(x)(\frac{\omega}{c})^2$
where $\delta\epsilon(x)$ is the fluctuating part of the dielectric constant.
On an ordered lattice with all wells the same depth, an electron is mobile for a range of energies.  However, in disordered media, well depths become more random and electrons with sufficient negative energy, may get trapped in regions where the random potential $V(x)$ is particularly deep \cite{Milovanov2002,Milovanov2001}.  The ability of electrons to tunnel out of the potential well depends on the probability of finding nearby potential fluctuations into which the trapped electron can tunnel. 

If we consider $V(x)$ to have a root-mean-square amplitude $V_{rms}$ and a length scale $a$ (the minimum scale of the network which relates to size of the basic elements or “conducting links” constituting the structure), on which random fluctuations in the potential take place, $E_a$ is the conduction band width which is defined by an energy scale $E_a\equiv{{\hbar^2}/{(2ma^2)}}$.  At the weak disorder limit, $V_{rms}\ll{E_a}$, a transition takes place at a critical energy ${E_c}\simeq{-V_{rms} 2/E_a}$.  Successive tunnelling events allow electrons of energy $>E_c$ to traverse the entire solid in a diffusive process leading to conductivity, whilst, electrons with energy $<E_c$ are trapped and do not conduct.  For energies $\gg{E_c}$, the electron traverses the solid with relative ease, provided disorder is below a critical level.  However, unlike conventional conduction, e-pairs in SC are additionally constrained by thermal instability above $T_c$.

In the case of HTSC, below $T_c$, the quantum tunnelling length for an electron pair is given by $l\sim{\hbar\sqrt{4mT}}$, where $T$ stands for the height of the barrier.  A superconducting state may therefore exist if $l\gtrsim{a}$ according to the relation 
\beq
T\lesssim\hbar^2/4ma^2
\label{eq.2}
\eeq
By contrast with electrons, a binding potential well does not exist for light.  However in a second localization mechanism originally described by Anderson \cite{Anderson1958}, photons (and electrons) can still be trapped in disordered networks.  At the weak disorder limit, when a quantum particle is inserted in a disordered system it will start to spread in a diffusive process, the wave being backscattered by impurities leading to weak localization effects.  The multiple scattering of the wave can enhance these perturbative effects to such a degree that they become spatially localized.  Below a critical level of disorder, there is a finite probability for the particle to return to the point at which it was inserted.  At very high levels of disorder, states become exponentially extended and this probability moves to zero.  However, due to the lack of a binding potential, photons have the ability to theoretically escape from disordered media, although in practice they can be trapped for extended periods of time \cite{Apalkov2002}.

CRL emerges under very specific conditions as the degree of disorder increases \cite{Apalkov2002}.  In the absence of mirrors, which are normally required to support coherent lasing, the disordered medium itself somehow takes on this role. For the disordered medium to play the role of a Fabry-Perot resonator, it is necessary that certain Eigen functions are completely, or almost localized. An almost localized solution can be viewed as a very high local maximum of the extended Eigen function $\psi(x)$.  If this maximum is viewed as a core, then the delocalized tail can be viewed as a source of leakage. The higher the local maximum of  $\psi(x)$, associated with increasing levels of disorder, the longer the period of localization.  At low levels of gain, leakage is not observable.  However, as gain increases this changes, with phase coherent photons escaping at multiple locations, observed as random lazing.

We conclude that established theory on localization in disordered networks appears to conflict with HTSC and CRL, which are supported by high levels of disorder. A new theory is required to address this specific issue.  Alpakov \cite{Apalkov2002} suggested that the link to the Fabry-Perot resonator could be identified with ‘disorder-induced resonance’ leading to macroscopic quantum coherence.  Milovanov and Rasmussen \cite{Milovanov2002} hint at a mechanism whereby the complex microscopic texture of heterogenous HTSC materials support “fractional harmonic modes” that could lead to macroscopic coherence.  In both HTSC and CRL, a common theme of macroscopic resonance in disordered networks emerges which we consider in section 3.

\subsection*{\centering{2.2. Fractal networks}}
In considering fractal structures we define three specific dimensions.  The dimension of the embedding Euclidean space $E^D$ where $D$ is the integer dimension, the fractal (Hausdorff) dimension $D_F$ and the spectral (Fracton) dimension $D_{fr}$.  Fractons are localized quantum oscillations of the local disorder in the system in a manner similar to extended modes (e.g. phonons) found in an ordered lattice \cite{Buttner1987,Milovanov2002,Bak2008,Courtens1989,Prester1999,Prester2001}. 

When the concentration $q$ of a fractal web embedded in $E^D$ goes to zero $(q\to0)$, there are no interconnecting conducting elements and zero conduction.  Conversely, as $q\to1$, a fractal web densely fills the Euclidean space as a percolating, multiscale, infinitely connected web.  At some point on the continuum from $0\to1$,  $q$ reaches a minimum critical concentration $q_c$.  At this 'percolation threshold', we have an infinite connected fractal web occupying a fraction of $E^D$, which conducts on large scales \cite{Milovanov2002}.  At some point $>q_c$ we expect to see the emergence of disorder induced localization effects.

\subsection*{\centering{2.3. Electron-pair coupling mechanisms}}

B\"uttner and Blumen \cite{Buttner1987} were amongst the first to suggest that HTSC could be supported by a partial (or total) fractal lattice with e-pair coupling being mediated by fractal phonon ($fr_{ph}$) modes which start to dominate above the frequency of the phonon to fracton crossover \cite{Courtens1989,Nakayama1994}.  This mechanism is analogous to phonon coupling in the BCS theory.  Their work, (supported by McMillan \cite{McMillan1968}) identified fracton density of states $\rho_{fr}$ (which determines the extent of coupling), and fracton frequency $\omega_{fr}$ as critical parameters influencing $T_c$.  The outcome of their analysis indicated a substantial increase in fracton mediated electron coupling strength in fractal structures, compared to that of phonons in conventional translation-invariant structures, that could account for e-pair coupling in HTSC.  Whilst this initial work, along with that of others \cite{Bak2008,Moulopoulos1995} offers insights into a possible mechanism, it appears incomplete as a theory to explain later experimental observations \cite{Hufner2008}, which suggests a more complex multicomponent mechanism at work.  In addition there have been concerns around a phonon coupled e-pair mechanism.  Many workers have ruled out this as an option due to the general lack of an isotope effect, although this observation is not universal \cite{Gweon2004} and a possible role for phonons in e-pair coupling is still under consideration \cite{Lanzara2001,Ohkawa2004,Citro2006}.

Copper oxides have become one of the most studied classes of HTSC materials.  Doping of the $p$-type cuprates with holes (positive charge carriers) distinguishes them from conventional superconductors \cite{Hufner2008}.  Electron orbitals in a uniform undoped parent compound, of two dimensional square $CuO_2$ planes, with lattice parameter $a = 3.8\mathring{A}$, between the nearest-neighbour Cu ions \cite{Tranquada2004} give rise to strong electron correlations leading to an antiferromagnetic (AF) insulator described by the Hubbard model.  Superconductivity (described by the Bose-Hubbard model), appears with the introduction of dopants into the insulating state, which coexist with AF fluctuations \cite{Tranquada2004}.  

Subsequent to B\"uttner and Blumen \cite{Buttner1987}, a number of analogous quasiparticle/excitation theories have been introduced, offering potential mechanisms for e-pair coupling.  The most widely reported are (AF) spin wave fluctuations (`magnons'), which are generally considered to play an important role in cuprate superconductors \cite{Monthoux1992,Prester1999,Prester2001,Moulopoulos1995,Arrigoni2004,Byczuk2010,Norman2005,Pickett2006}.  Magnetic moments can fluctuate about their average orientations and can theoretically be influenced by dopant ordering.  Pickett \cite{Pickett2006} indicated that dopant concentration as low as $x\approx 0.03$ is sufficient to destroy long range magnetic order.  Prester \cite{Prester1999,Prester2001} suggested that dopants introduce random obstacles for spin waves, inducing localized fracton modes, which we will refer to in what follows as fractal magnons ($fr_{mg}$).  He proposed that an $fr_{mg}$ mediated coupling mechanism is responsible for e-pairing in HTSC.

It has been proposed by a number of workers that dopants self-organize into filamentous stripes \cite{Fratini2010,Milovanov2002,Prester1999,Prester2001} creating a fractal structure that alternates with antiferromagnetic (insulating) regions within copper-oxide planes created by Coulomb repulsion \cite{Byczuk2010,Norman2005}. This is mirrored in at least one study by a fractal distribution of local lattice distortions which appear to be influenced by dopant order \cite{Poccia2012}. In addition the physical (granular) micro structure of the material may also have an influence on the distribution of oxygen dopants \cite{Buttner1987}.  This seems highly plausible, as we shall discuss in section 3.  In a separate study on three different copper oxide superconducting materials \cite{Chadzynski2008}, the granular structure of the different materials was found to vary with surface fractal dimensions ranging from 2.3 to 2.7, suggesting the potential to influence this specific variable.  

When considering dopant disorder on the micrometer scale, Fratini {\it{et al}} \cite{Fratini2010} found annealing temperatures in thermal treatments had a significant influence on dopant self-assembly at constant doping levels.  In a case study, temperatures ranging from $180K$ to $310K$ resulted in a continuous order to disorder transition, with disorder peaking at around $310K$, after which it declined to a continuous (homogenous) phase at around $350K$.

In additional studies at constant doping \cite{Fratini2010}, the correlation length associated with the fractal dopant network (described by a spatial intensity correlation function) increased with increasing dopant levels, varying between 50 and $400\mu{m}$ depending on annealing conditions.  Interestingly, $T_c$ was observed to increase as values for the spatial intensity correlation function increased.  These observations offer important insights into our understanding of the emergence of macroscopic quantum coherence in fractal networks, which on face value appears to be in conflict with the theory of localization in disordered networks \cite{Milovanov2002,Anderson1958}.  

Bak \cite{Bak2008} noted `one expects degradation of the SC state with increasing $D_F$, but that there appears to be a mechanism which can compensate this destructive effect'.  This suggests only partially localized DOS (LDOS), with tunnelling between LDOS facilitated by an unidentified mechanism, which strongly relates to the scale free (fractal) ordering of the dopants.  We consider the identification of this mechanism as a key component in explaining HTSC which we discuss in sections 3.2.4 and 3.2.5.

\subsection*{\centering{2.4. The Pseudogap and its relations with HTSC}}

One of the most noted properties of HTSC materials is the presence of two distinct electron coupled energy gaps \cite{Hufner2008,Kohsaka2008}.  The first relates directly to the energy gap associated with the SC phase ($T_c$). The second is a higher energy pseudogap (PG),  represented by $T^*$ in what follows, which is characterised by LDOS, which on the face of it should not contribute to conduction on the macroscale.  An issue that we consider in more detail in section 3.2.4.

\begin{figure}[!ht]
\begin{center}
\includegraphics[width=8.6cm]{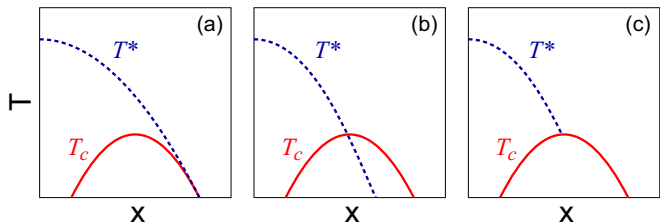} %
\caption{\small{Alternative phase diagrams for the pseudogap and superconducting gap.}}
\label{fig1}
\end{center}
\end{figure}

In a comprehensive review, H\"ufner {\it{et al}} \cite{Hufner2008} reported that the extent of the difference between $T_c$ and $T^*$ is directly related to the level of hole doping (x) which has been described in three different phase diagrams (Figure~\ref{fig1}) reproduced from the paper.  There is some debate around which phase diagram is correct \cite{Hufner2008}. 

\begin{figure}[!ht]
\begin{center}
\includegraphics[width=8.6cm]{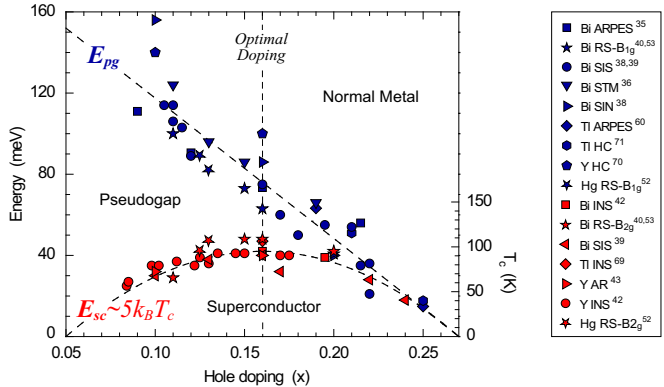} %
\caption{\small{Detailed phase diagram of Figure 1(a)}.}
\label{fig2}
\end{center}
\end{figure}

Figure~\ref{fig2} (reproduced from H\"ufner {\it{et al}} \cite{Hufner2008}) offers a more detailed picture of Figure~\ref{fig1}(a) for the SC and PG phase, relating $T_c$ and the respective energy gaps ($E_{sc}$ and $E_{pg}$) with dopant levels.  As indicated in the key to Figure~\ref{fig2}, data are collated from a number of different process methodologies and materials, suggesting the phase relations are reasonably robust, although multiple data sources does add variability, with value predictions being inevitably less precise.  The precise cutoff for conduction at low levels of doping is not clear on this figure.  However, below specific levels of doping the material becomes an AF insulator \cite{Norman2010}. 

Looking at the overall picture represented by Figure~\ref{fig2}, the PG phase is expressed by a linear relationship which declines with increasing dopant level, whilst the SC phase follows a parabola which peaks at optimum doping ($x\sim 0.15-0.17$).  

At low (sub optimum) doping values ($x\sim0.08$), the energy level $E_{sc}$ in the superconducting phase  is at a minimum whilst that in the pseudogap $E_{pg}$ is at its maximum.  The phase diagram does not specifically show values for $E_{pg}$ at this level.  However, a rough extrapolation of relations for the  $E_{pg}$  indicates that values for $T^*$ as high as 350K ($77^0C$) may be possible at low doping levels.  This is in line with independently reported values \cite{Colossal2014}.  Irrespective of the highest values that may be achieveable, it is an important result.  It indicates that electron pairs can exist above room temperature under the right conditions, suggesting the possibility of higher values for $T_c$ in the SC phase, if we can better understand the conditions required to facilitate tunnelling between PG LDOS and the SC condensate. 

At doping levels beyond the optimum, $T_c$ decreases from its maxima, whilst $T^*$ continues to drop, with both values converging at a doping level of $x\sim0.22$.  H\"ufner {\it{et al}} \cite{Hufner2008} commented that it is not yet clear if the PG and the SC phase coexist as a coherent condensate at this level.  However, they suggest that the two phases may remain distinctly different with different tunnelling mechanisms.  Extrapolation of the data in figure 2 with increasing doping levels (beyond x$\sim0.25$) suggests a reversion to conventional SC or its possible disappearance.

Phase relations represented by the different figures offer a good review of experimental observations.  However, a theory to explain the results has not yet been elucidated.  We revisit these issues in sections 3.2.4 and 3.2.5, once the theoretical framework has been outlined.

McElroy {\it{et al}} \cite{McElroy2005} identified a key challenge to determine how dopants influence the superconducting electronic structure at the atomic scale.  They suggested that dopant atoms may have two diametrically opposing influences on HTSC, being destructive at the atomic scale but supportive globally (via enhanced carrier densities). 

As observed by Byczuk {\it{et al}} \cite{Byczuk2010}, it might reasonably be expected that since electronic (AF) interactions and disorder can both (and separately) induce a metal insulator transition, their simultaneous presence might collectively act to increase localization rather than lead to HTSC.  However, the interplay between disorder and interactions leads to subtle many body effects \cite{Byczuk2010}.  Neutron scattering studies \cite{Tranquada2004} on stripe ordered copper oxides suggests that quantum magnetic fluctuations linked to charge disorder, possibly dynamic in nature, play an essential role in HTSC. In the HTSC cuprates, Byczuk {\it{et al}} \cite{Byczuk2010}, reports a synergistic interplay between the AF insulator and dopant induced disorder with a clearly defined peak associated with optimal doping, which correlates with maximum $T_c$.

\subsection*{\centering{3. A new approach to HTSC}}
\subsection*{\centering{3.1. Challenges}}

A number of theories and extensive data contribute to our understanding of HTSC. However, so far no single theory has been universally accepted as a satisfactory explanation of experimental observations.  We postulate that this lack of consensus relates to the complex interaction of a number of mechanisms that require a more fundamental consideration and a better understanding of their interrelations.  The complexity is increased by number of different possible quasiparticle based electron coupling mechanisms which appear to be dependent on the material being considered. 

To obtain a clearer picture, we need a deeper understanding of a number of key issues which we have identified.  These include the following:

\begin{itemize}
\item Support or refute alternative e-pair coupling mechanisms that have been proposed in the literature to date \cite{Buttner1987,Prester1999,Prester2001,Ma2014,Nottale2014}.
\item Resolve the proposed dual e-pair coupling mechanisms responsible for the PG and SC phase \cite{Hufner2008}. 
\item Identify the correct phase diagram with respect to PG relations shown in Figure~\ref{fig1} and Figure~\ref{fig2} \cite{Hufner2008}.
\item Identify the role of the PG in HTSC \cite{Hufner2008}.
\item Identify the role of dopants in HTSC \cite{McElroy2005}.
\item Address e-pair localization theory in disordered networks \cite{Milovanov2002,Anderson1958}.
\item Identify the mechanism whereby an expected diffusive process in a fractal network leads to macroscopic quantum coherence in both CRL and HTSC \cite{Apalkov2002,Milovanov2002,Bak2008}.  
\item Identify relations between the AF insulator and HTSC \cite{Tranquada2004}.
\end{itemize}

\subsection*{\centering{3.2. Theoretical framework}}
\subsection*{\centering{ {\it{3.2.1.Introduction}}}}

In what follows we construct a theoretical approach to address the challenges identified in section 3.1.   A key component of this work is based upon concepts and a mathematical approach established in the theory of Scale Relativity \cite{Nottale2014,Auffray2008,Nottale1993,Nottale2008,Nottale2009,Nottale2011}.  The theory is founded on the principle that the apparently smooth geodesics of space-time at the macro-scale described by general relativity are an incomplete description of the structure of space-time at the micro-scale.  If we are to understand quantum mechanics in terms of space-time geometry then we need to rethink its fundamental structure in a way that reflects our understanding of the mechanics.  In simple terms, the hypothesis states that the structure of space has both a smooth (differentiable) component at the macro-scale and a chaotic, fractal (non-differentiable) component at the micro-scale.  At the macroscale, the fractal component and its influence is small and generally considered unimportant in classical physics.  However, at the microscale, the fractal component and its influence dominate, with quantum laws originating in the underlying fractal geometry of space-time, the transition taking place at the de Broglie length scale.  The origins of this hypothesis can be traced back to Feynman \cite{Feynman1965}, who suggested that the typical quantum mechanical paths that are the main contributors to the 'path integral', are infinite, non differentiable and fractal (to use current terminology).  

In the theory, the transition of a system from the classical to the quantum regime only becomes effective when three key conditions have been fulfilled.  The first is that the paths or trajectories are infinite in number, leading to a non-deterministic and probabilistic, fluid like description in which the velocity $v(t,dt)$ on a particular geodesic is replaced by a velocity field $v[x (t),t]$ in which the concept of a single trajectory has no meaning. The second, that the trajectories are fractal curves which leads to a fractal velocity field $V=[x(t,dt),t,dt]$. The velocity field is therefore defined as a fractal function, explicitly dependent on resolutions and divergent when the scale interval tends to zero (this divergence is the manifestation of non-differentiability).  In the final condition non-differentiability leads to two fractal velocity fields $V_+[x(t,dt),t,dt]$ and $V_-[x(t,dt),t,dt]$, which are no longer invariant under transformation ${|dt|}\to-{|dt|}$ in the non-differentiable case. 

These two fractal velocity fields may in turn be decomposed i.e.
\begin{eqnarray}
V_+=v_+[x(t),t]+w_+[x(t,dt),t,dt],\\ 
V_-=v_-[x(t),t]+w_-[x(t,dt),t,dt].
\end{eqnarray}
The (+) and (-) velocity fields comprise a `classical part' $(v_+,v_-)$ which is differentiable and independent of resolution, and of fractal fluctuations of zero mean $(w_+, w_-)$, explicity dependent on the resolution interval $dt$ and divergent at the limit $dt\to0$.  A simple and natural way to account for this doubling consists in using complex numbers and the complex product \cite{Nottale2011}.  The three properties of motion in a fractal space lead to a description of a geodesic velocity field in terms of a complex fractal function.  The full complex velocity field reads
\beq
\widetilde{V}={\widehat V} + {\widehat W} = \left(\frac{v_{+}+v_{-}}{2} -i \, \frac{v_{+}-v_{-}}{2} \right) + \left(\frac{w_{+}+w_{-}}{2}-i \, \frac{w_{+}-w_{-}}{2}\right).
\label{eq.5}
\eeq
In this interpretation, the jump from a real to a complex description is the origin of the real and imaginary components in the wave function.  However, as we show in what follows this is not constrained to the microscale.

\subsection*{\centering{{\it{3.2.2. A geodesic approach to quantum mechanics}}}}
If we consider elementary displacements along these geodesics, this reads $dX_\pm=d_\pm x+d\xi_\pm$, where (in the case of a critical fractal dimension $D_F=2$ for the geodesics in the case of standard quantum mechanics) 
\beq
d_{\pm} x= v_{\pm} \; dt, \;\;\;
d\xi_{\pm}=\zeta_{\pm} \, \sqrt{2 \widetilde{D}}  \, |dt|^{1/2}.
\label{eq.6}
\eeq
$d\xi$ represents the fractal fluctuations or fractal part of the displacement $dX$. This interpretation corresponds to a Markov-like situation of loss of information from one point to another, without correlation or anti-correlation. Here $\zeta_{\pm}$ represents a purely mathematical dimensionless stochastic variable such that $\langle\zeta_{\pm}\rangle=0$ and $\langle\zeta_{\pm}^2\rangle=1$, the mean $\langle \rangle$ being described by its probability distribution.  $\widetilde{D}$ is a fundamental parameter which characterizes the amplitude of fractal fluctuations.  Since $d\xi$ is a length and $dt$ a time, it is given by the relation
\beq
{\widetilde D}= \frac{1}{2}\, \frac{\langle\enspace \!\!d \xi^2 \!\!\enspace\rangle}{dt}.
\label{eq.7}
\eeq
When considering the geodesics of a fractal space, the real and imaginary parts of the velocity field can be expressed in this case in terms of the complex velocity field\footnote{We note that the full complex velocity field $\widetilde{V}={\widehat V} + {\widehat W}$  introduced in Eq. (5) also includes the divergent part of zero mean ${\widehat W}$ from which a quantum mechanical particle and its properties can be derived.  For simpliity ${\widehat W}$ is not considered here.  For a detailed analysis of the full velocity field  $\widetilde{V}$ and its implications outside of the current work see \cite{Nottale2011}.}
\beq
{\widehat V}= V- i U.
\label{eq.cvfV}
\eeq
This equation captures the essence of the principle of relativity in which any motion, however complicated and intricate the path, should disappear in the proper reference system
\beq
{\widehat V}= 0.
\label{eq.V0}
\eeq
We now introduce the complex `covariant' derivative operator $\widehat{d}/dt$, which includes the terms which allow us to recover differentiable time reversibility in terms of the new complex process \cite{Nottale1993,Nottale2011} 
\beq
\frac{\widehat{d}}{dt} = \frac{1}{2} \left(\frac{d_+}{dt}+ \frac{d_-}{dt}\right)-\frac{i}{2}\left(\frac{d_+}{dt} - \frac{d_-}{dt}\right).
\label{eq.10}
\eeq
Applying this operator to the position vector yields the differentiable part of the complex velocity field 
\beq
{\widehat V} = \frac{\widehat{d}}{dt} \,x = V -i U = \frac{v_+ + v_-}{2} - i 
\;\frac{v_+ - v_-}{2}.
\label{eq.11}
\eeq
Deriving Eq.~(\ref{eq.V0}) with respect to time, it takes the form of a free inertial equation devoid of any force
\beq
\frac{\widehat{d}}{dt} \,{\widehat V} =0.
\label{eq.12}
\eeq
In the case of a fractal space, the various effects can be combined in the form of a total derivative operator \cite{Nottale1993,Nottale2011},
\beq
\frac{\widehat{d}}{\partial t}=\frac{\partial}{\partial t} + {\widehat V} . {\nabla} -i {\widetilde D} \Delta.
\label{eq.totd}
\eeq
The fundamental equation of dynamics becomes, in terms of this operator
\beq
m \, \frac{\widehat{d}}{dt} \,{\widehat V} =- \nabla \phi
\label{eq.dyn}
\eeq
which is now written in terms of complex variables and of the complex time derivative operator.  Which, in the absence of an exterior field $\phi$, is a geodesic equation.

\subsection*{{\it{3.2.2.1. From Newton to the Schr\"odinger equation}}}

After expansion of the covariant derivative, the free-form motion equations of general relativity can be transformed into a Newtonian equation in which a generalized force emerges, of which the Newton gravitational force is an approximation.  In an analogous way, the covariance induced by scale effects leads to a transformation of the equation of motion, which, as we shall demonstrate through a number of steps, becomes after integration, the Schr\"odinger equation.  In the construction of this approach we note that whilst equations take a classical form, this form is applied to non differentiable geometry, so that the result is no longer classical

\subsection*{\it{Momentum}}

Due to the complex nature of the velocity field $\widehat V$, the classic equation $p=mv,$ can be generalized to its complex representation \cite{Nottale2011}.  
\beq
{\widehat P} = m {\widehat V},
\label{eq.15}
\eeq
so that the complex velocity field $\widehat V$ is potential (irrotational), given by the gradient of the complex action,
\beq
{\widehat V} = \frac{ \nabla {\widehat S}}{m}.
\label{eq.cvf}
\eeq
We now introduce a complex function $\psi$ identifiable with a wave function or state function, which is another expression for the complex action $\widehat S$,
\beq
\psi = e^{i{\widehat S}/S_0}.
\label{eq.psi}
\eeq
The factor $S_0$ has the dimension of an action, i.e. of an angular momentum with $S_0=\hbar$ in the case of standard quantum mechanics (QM), where $\hbar$ is a geometric property of the fractal space, defined through the fractal fluctuations as $\hbar=2m{\widetilde D}=m\langle{d\xi^2}\rangle/dt$.

The next step consists of making a change of variable in which one connects the complex velocity field Eq.~(\ref{eq.cvf}), to a wave function, $\psi$ where $\ln\psi$ plays the role of a velocity potential according to the relation
\beq
{\widehat V} = - i \, \frac{S_0}{m} \, \nabla (\ln \psi).
\label{eq.vfwf}
\eeq
The complex momentum Eq.~(\ref{eq.15}) may now be written under the form
\beq
{\widehat P}= - i \,S_0 \, \nabla (\ln \psi),
\label{eq.19}
\eeq
i.e. 
\beq
{\widehat P} \psi= - i \,S_0 \, \nabla \psi.
\label{eq.20}
\eeq
In the case of standard quantum mechanics ($S_0=\hbar$), this relation reads ${\widehat P} \psi= - i \,\hbar \, \nabla \psi$, i.e. its a derivation of the principle of correspondence for momentum, $p \to -i\, \hbar\, \nabla$ where the real part of the complex momentum $\widehat P$ is, in the classical limit, the classical momentum $p$.  The `correspondence' is therefore understood as between the real part of a complex quantity and an operator acting on the function $\psi$.  However, thanks to the introduction of the complex momentum of the geodesic fluid, it is no longer a mere correspondence, it has become a genuine equality.
The same follows for angular momentum $L=rp$, which can also be generalized to the complex representation 
\beq
{\widehat L} \psi =  - i \,S_0 \; r \times \nabla \psi,
\label{eq.21}
\eeq
so that we recover, in the standard quantum case $S_0=\hbar$, the correspondence principle for angular momentum, which again emerges as an equality.

\subsection*{\it{Remarkable identity}}

We now write the fundamental equation of dynamics Eq.~(\ref{eq.dyn}) in terms of the new quantity $\psi$. 
\beq
i S_0 \frac{\widehat{d}}{dt}(\nabla \ln \psi) = \nabla \phi.
\label{eq.22}
\eeq
We note that $\widehat{d}$ and $\nabla$ do not commute.  However, as we shall see in what follows ${\widehat{d}}(\nabla \ln \psi)/dt$, is a gradient in the general case.
  
Replacing $\widehat{d}/{dt}$  by its expression, given by Eq.~(\ref{eq.totd}), yields
\beq
\nabla   \phi  =  i S_0 \left(\frac{\partial}{\partial t} + {\widehat V}. 
\nabla - i {\widetilde D} \Delta\right) (\nabla \ln \psi),
\label{eq.23}   
\eeq
and replacing once again ${\widehat V}$ by its expression in Eq.~(\ref{eq.vfwf}), we obtain 
\beq
\nabla   \phi  =   i S_0 \left\{ \frac{\partial }{\partial t} \nabla   
\ln\psi   - i \left[  \frac{S_0}{m} (\nabla   \ln\psi  . \nabla   )
(\nabla   \ln\psi ) + {\widetilde D} \Delta (\nabla   \ln\psi )\right]\right\} .
\label{eq.nablaphi}
\eeq
Consider now the identity \cite{Nottale1993}  
\beq
(\nabla \ln f)^{2} + \Delta \ln f =\frac{\Delta f}{f} \; ,
\label{eq.25}
\eeq
which proceeds from the following tensorial derivation
\begin{eqnarray}
\partial_{\mu} \partial^{\mu} \ln f +\partial_{\mu} \ln f \partial^{\mu} 
\ln f &=& \partial_{\mu} \frac{\partial^{\mu} f}{f}+\frac{\partial_{\mu} f}
{f}\frac{\partial^{\mu} f}{f} \nonumber \\
 &=& \frac{f \partial_{\mu} \partial^{\mu} f - 
\partial_{\mu} f \partial^{\mu} f}{f^{2}}+\frac{ \partial_{\mu} f 
\partial^{\mu} f}{f^{2}} \nonumber \\
&=& \frac{\partial_{\mu} \partial^{\mu} f}{f} \; . 
\label{eq.26}
\end{eqnarray}
When we apply this identity to $\psi$ and take its gradient, we obtain
\beq
\nabla\left(\frac{\Delta \psi}{\psi}\right)=\nabla[(\nabla \ln \psi)^{2} + 
\Delta \ln \psi] .
\label{eq.27}
\eeq
The second term on the right-hand side of this expression can be transformed, 
using the fact that $\nabla$ and $\Delta$ commute, i.e.,
\beq
\nabla \Delta =\Delta \nabla . 
\label{eq.28}
\eeq
The first term can also be transformed thanks to another identity,
\beq
\nabla (\nabla f)^{2}=2 (\nabla f . \nabla) (\nabla f) ,
\label{eq.29}
\end{equation}
that we apply to $f=\ln \psi$. We finally obtain \cite{Nottale1993}
\beq
\nabla\left(\frac{\Delta \psi}{\psi}\right)= 2 (\nabla \ln\psi . \nabla )
(\nabla \ln \psi )  + \Delta (\nabla \ln \psi).
\label{eq.30}
\eeq
This identity can be still generalized thanks to the fact that $\psi$ appears only through its logarithm in the right-hand side of the above equation. By replacing  $\psi$ with $\psi^{\alpha}$, we obtain the general remarkable identity \cite{Nottale2008b}
\beq
\frac{1}{\alpha} \; \nabla\left(\frac{\Delta  \psi^{\alpha}}{\psi ^{\alpha}}\right)= 2\alpha \, (\nabla \ln \psi . \nabla )(\nabla \ln \psi) + \Delta (\nabla \ln \psi).
\label{eq.RI}
\eeq

\subsection*{\it{Schr\"odinger equation}}
We recognize in the right-hand side of Eq.~(\ref{eq.RI}) the two terms of Eq.~(\ref{eq.nablaphi}), which were respectively in factor of $S_0$ and $\widetilde D$.  Therefore, by writing the above remarkable identity in the case
\beq
\alpha= \frac{S_0}{2 m {\widetilde D}},
\label{eq.alpha}
\eeq
the whole motion equation becomes a gradient,
\beq
\nabla   \phi  =  2m  {\widetilde D}\left[ i\frac{\partial }{\partial t} \nabla   
\ln\psi^\alpha  +  {\widetilde D}   \nabla\left(\frac{\Delta  \psi^{\alpha}}{\psi ^{\alpha}} \right)\right],
\label{eq.33}
\eeq
and it can therefore be generally integrated, in terms of the new function
\beq
\psi^{\alpha}=  (e^{i{\widehat S}/S_0})^\alpha=e^{i{\widehat S}/2 m {\widetilde D}}.
\label{eq.34}
\eeq
which is more general than in standard QM, for which $S_0=\hbar=2m\widetilde D$.  Eq.~(\ref{eq.alpha}) is actually a generalization of the Compton relation.  This means that the function $\psi$ becomes a wave function only provided it comes with a Compton-de Broglie relation, a result which is  naturally achieved here.  Without this relation, the equation of motion would remain of third order, with no general prime integral.

The simplification brought by this relation means that several complicated terms are compacted into a simple one and that the final remaining term is a gradient, which means that the fundamental equation of dynamics can now be integrated in a universal way.  The function $\psi$ in Eq.~(\ref{eq.psi}) is therefore finally defined as
\beq
\psi = e^{i{\widehat S}/2m{\widetilde D}},
\label{eq.35}
\eeq
which is a solution of the fundamental equation of dynamics, Eq.~(\ref{eq.dyn}), 
which now takes the form
\beq
\frac{\widehat{d}}{dt} {\widehat V} = -2 {\widetilde D} \nabla \left(i \frac{\partial}
{\partial t} \ln \psi + {\widetilde D} \frac{\Delta \psi}{\psi}\right) = 
-\frac{\nabla \phi}{m}.
\label{eq.36}
\eeq	
Using the fact that $d ln\psi=d\psi/\psi$, the full equation becomes a gradient,
\beq
\nabla\left[\frac{\phi}{m}-2 {\widetilde D} \nabla \left(\frac{i \partial\psi/\partial t+\widetilde D\Delta\psi}{\psi}\right)\right] = 0.
\label{eq.37}
\eeq	
Integrating this equation finally yields a generalized Schr\"odinger equation,
\beq
{\widetilde D}^2 \Delta \psi + i {\widetilde D} \frac{\partial}{\partial t} \psi - 
\frac{\phi}{2m}\psi = 0,
\label{eq.38}
\eeq
where $\widetilde D$ identifies with the amplitude of the quantum force, but becomes more generalized than its standard QM equivalent $(\hbar/2m)$, accommodating both the one body and many body case (either distinguishable or indistinguishable particles), as well as the possibility of macroscopic values.

\subsection*{{\it{3.2.2.2. Fluid representation with a macroscopic quantum potential}}}
We now demonstrate the fundamental meaning of the wave function as a wave of probability, and that the geodesic equation can take not only a Schr\"odinger form, but also a fluid dynamics form with an added quantum potential.  Again we use the constant  $\widetilde D$ instead of the standard QM expression $\hbar/2m$.  We begin by writing the wave function under the form $\psi=\sqrt{P}\times{e}^{iA/\hbar}$, decomposing it in terms of a phase, defined as a dimensioned action $A$ and a modulus $P= |\psi|^2$ , which gives the number density of virtual geodesics \cite{Nottale2011,Nottale2007}.  This function becomes naturally a density of probability.  The function $\psi$, being a solution of the Schr\"odinger equation and subjected to the Born postulate and to the Compton relation, therefore owns most of the properties of a wave function.

The complex velocity field $\widehat V$ Eq.~(\ref{eq.cvfV}) can be expressed in terms of the classical (real) part of the velocity field $V$ and of the number density of geodesics $P_N$, which is equivalent as we have seen above, to a probability density $P$ where
\beq
\Vsc= V- i \Ds \nabla \ln P.
\label{eq.39}
\eeq
The quantum covariant derivative operator thus reads
\beq
\frac{\dfr}{\d t}=\frac{\d}{\d t} + V. \nabla -i \Ds \: ( \nabla \ln P.\nabla+\Delta).
\label{eq.40}
\eeq
When we introduce an exterior scalar potential $\phi$, the fundamental equation of dynamics becomes
\beq
\l( \frac{\d}{\d t} + V. \nabla -i \Ds \: ( \nabla \ln P.\nabla+\Delta) \r) ( V- i \Ds \nabla \ln P)=-\frac{ \nabla \phi}{m}.
\label{eq.41}
\eeq
The imaginary part of this equation,
\beq
\Ds \: \l[  ( \nabla \ln P.\nabla+\Delta)  V + \l( \frac{\d}{\d t} + V. \nabla\r)\nabla \ln P \r]=0,
\label{eq.42}
\eeq
takes, after some calculations, the following form
\beq
\nabla\l[   \frac{1}{P}  \l( \frac{\d P}{\d t} + \text{div} ( P V) \r) \r]=0,
\label{eq.43}
\eeq
which can finally be integrated in terms of a continuity equation:
\beq
\frac{\d P}{\d t} + \text{div} ( P V)=0.
\label{eq.cont}
\eeq
The real part,
\beq
\l( \frac{\d}{\d t} + V. \nabla\r) V=-\frac{ \nabla \phi}{m} +\Ds^2 \: ( \nabla \ln P.\nabla+\Delta) \nabla \ln P,
\label{eq.45}
\eeq
takes the form of a Euler equation, 
\beq
m \l( \frac{\d}{\d t} + V. \nabla\r) V=- \nabla \phi +2 m \Ds^2\nabla \l( \frac{\Delta\sqrt{P}}{\sqrt{P}}\r),
\label{eq.Euler}
\eeq
which describes a fluid subjected to an additional quantum potential $Q$ that depends on the probability density $P$
\beq
Q=-2 m \Ds^2 \: \frac{ \Delta \sqrt{P}}{\sqrt{P}}.
\label{eq.Q}
\eeq

The obtention of Eq's (\ref{eq.cont}) and (\ref{eq.Euler}) definitively prove the identification of $P$ with a density of probability.  $Q$ is a manifestation of the fractal geometry and probability density, with fractal fluctuations of space-time (at the micro-scale) leading to the emergence of a `fractal field', a potential energy (`quantum potential') and a quantum force which are directly analogous with the geometric origins of the gravitational field, gravitational potential and gravitational force which emerge as a manifestation of the curved geometry of space-time.  The potential energy $Q$ is implicitly contained in the Schr\"odinger form of the equation as the standard quantum potential, but here established from the geodesic equation as a fundamental manifestation of the fractal geometry.  It is only made explicit when reverting to the Euler representation.   We note that the approach followed is similar to the Madelung transformation \cite{Madelung1927}. However, since the Schr\"odinger equation is obtained as a reformulation of the geodesic equation, it is possible to go directly from the covariant equation of dynamics Eq.~(\ref{eq.dyn}) to the fluid mechanics equations without defining the wave function or passing through the Schr\"odinger equation. 

This new geometric approach to quantum theory offers some important new insights.  Classical quantities relate to the quantum world as averages.  Conversely quantum properties remain at the heart of the classical world.  We have shown that the action of fractality and irreversibility on small time scales can manifest itself through the emergence of a macroscopic quantum-type potential energy, in addition to the standard classical energy balance.  This potential energy leads to the possibility of macroscopic quantum effects, which are normally masked by classical motion, but can be observed given the right conditions.  Examples include Bose Einstein condensates, superfluids and superconductivity when the de Broglie length scale becomes macroscopic as the momentum of particles decrease at very low temperatures.  However, as we shall show, HTSC cannot be simply explained in terms of an extended $\lambda_{deB}$ due the higher temperatures involved, which requires an account of e-pair instability at higher temperatures.  In addition, we have to account for localization effects in disordered systems.

\subsection*{{\it{3.2.2.3. Ginzburg-Landau equation}}}
Relations between the proposed theory and SC can be found in its phenomenological Ginzburg-Landau equation. In an analogy to Eq.~(\ref{eq.Euler}), it is possible to recover such a non-linear Schr\"odinger equation simply by adding a quantum-like potential energy to a standard fluid including a pressure term \cite{Nottale2009}.  We do this by replacing probability density $P$ with matter density $\rho$, as the velocity field $V$ of a fluid in potential motion can also be combined in terms of a complex function $\psi = \sqrt{\rho} \times e^{{i A/\hbar}}$ \cite{Nottale2009}.  In what follows, since $\rho$=$PM$ (where $M$ is the total mass of the fluid in the volume considered), $m = 1$.

Consider a Euler equation with a pressure term and a quantum potential term
\begin{equation}
\l(\frac{\partial}{\partial t} + V \cdot \nabla\r) V  = -\nabla \phi-\frac{\nabla p}{\rho}+2{\widetilde D}^2 \,\nabla \l( \frac{\Delta \sqrt{\rho}}{\sqrt{\rho}}\r).
\label{eq.48}
\end{equation}
where ${\nabla p}/{\rho}=\nabla w$ is itself a gradient, which is the case of an isentropic fluid, and, more generally, of every case when there is a state equation which links $p$ and $\rho$, its combination with the continuity equation can be still integrated in terms of a Schr\"odinger-type equation \cite{Nottale1997},
\beq
{\Ds}^2 \Delta \psi + i {\Ds} \frac{\partial}{\partial t} \psi - \frac{\phi+w}{2}\psi = 0.
\label{eq.49}
\eeq
In the acoustic approximation, the link between pressure and density writes $p-p_0=c_s^2(\rho-\rho_0)$, where $c_s$ is the speed of sound in the fluid, so that $\nabla p/\rho=c_s^2 \, \nabla \ln \rho$. Moreover, when $\rho-\rho_0 \ll \rho_0$, one may use the additional approximation $c_s^2 \, \nabla \ln \rho \approx (c_s^2 /\rho_0) \nabla \rho$, and the equation obtained takes the form of the  Ginzburg-Landau equation of superconductivity (here in the absence of a magnetic field),
\beq
{\Ds}^2 \Delta \psi + i {\Ds} \frac{\partial}{\partial t} \psi - \beta \, |\psi|^2 \, \psi= \frac{1}{2} \, \phi  \; \psi,
\label{eq.50}
\eeq
with $\beta={c_s^2}/{2 \rho_0}$. In the highly compressible case, the dominant pressure term is rather of the form  $p \propto \rho^2$, so that $p/\rho \propto \rho= |\psi|^2$, and one still obtains a non-linear Schr\"odinger equation of the same kind \cite{Noore1994}.

\subsection*{\centering{{\it{3.2.3. Proposed e-pair coupling mechanisms for the SC and PG phase}}}}

As a first step in constructing a theory of macroscopic coherence in HTSC we propose two separate, evidence based e-pair coupling mechanisms relating to the SC and PG phase.  In both scenarios we follow a standard quantum mechanics approach based on $\hbar$.  To do this the theoretical framework outlined in sections 3.2.1 to 3.2.2.2 is not strictly required.  However, to ensure a consistent approach within the paper and to show the natural links to the theory of scale relativity and later discussion of macroscopic quantum effects associated with HTSC, e-pair coupling is described within the context of the proposed theory, which addresses both microscopic and macroscopic scales.

\subsection*{{\it{3.2.3.1. The SC phase}}}

If we take the example of $p$-type cuprates, dopants have been reported as ordering on a scale free (fractal) network \cite{Fratini2010}, creating electronic disorder at the microscopic level.  In an example like $Bi_2Sr_2CaCu_2O_8+x$ \cite{Pan2001}, the local density of states (LDOS) creates ‘hills’ and ‘valleys’ of size $\approx 30\mathring{A}$, which strongly correlate with the energy gap $E_g$.  As previously reported \cite{McElroy2005,Prester1999,Prester2001,Nottale2014}, the minima of LDOS modulations preferentially occur at the dopant defects.  The regions with sharp coherence peaks, usually associated with strong superconductivity, are found to occur between the dopant clusters, near which the SC coherence peaks are suppressed.  

To understand this picture more clearly we are assisted by a previously unexplained phenonena.  Across a range of materials only $\approx 20\%$ of the total dopant induced charge is mobile and contributes to the superfluid at optimum doping \cite{Tanner1998}.  The $\approx 80\%$ static dopants have a significant impact on the topology of the cuprate surface.

During the annealing stage of the manufacturing process, dopants diffuse in a thermodynamically driven process (described in more detail in section 3.2.5) to create a macroscopic fractal network \cite{Fratini2010}.  At an individual level, a single dopant is repulsive (it can be seen as a quantum "pressure").  However, due to random fluctuations, some zones, devoid of dopants and surrounded by an increased number density of dopants are formed (with a Poisson distribution).  The combination of these repulsive external forces creates an attractive potential well, allowing the appearance of bound states, in a mechanism which we now describe.

We denote $\psi_n$ the wave function of all doping charges, $\psi_s$ the wave function of the fraction of carriers tied in e-pairs forming the superconducting fluid and $\psi_d$ the wave function of the fraction of charges, which do not participate in superconductivity, 
i.e. 
\beq
\psi_n=\displaystyle\sum\limits_{n=1}^N {\psi_{s_n}} +\displaystyle\sum\limits_{n=1}^N {\psi_{d_n}}
\label{eq.51}
\eeq
The doping induced charges constitutes a quantum fluid which is expected to be the solution of a Schr\"odinger equation 
\beq
\frac{\hbar^2}{2m} \Delta \psi_n + i \hbar \frac{ \d \psi_n}{\d t}= \phi\;  \psi_n,
\label{eq.52}
\eeq 
where $\phi$ is a possible external scalar potential, and where we have neglected the magnetic effects as a first step.

Separating the two contributions  $\psi_s$ and  $\psi_d$ in this equation. We obtain:
\beq
\frac{\hbar^2}{2m} \Delta \psi_s + i \hbar \frac{ \d \psi_s}{\d t}- \phi\;  \psi_s =-\frac{\hbar^2}{2m} \Delta \psi_d - i \hbar \frac{ \d \psi_d}{\d t}+\phi \; \psi_d.
\label{eq.psispsidschrodinger}
\eeq
We now introduce explicitly the probability density $\rho$ and phase, which we define as a dimensioned action $A$, of the wave functions  $\psi_s= \sqrt{\rho_s} \times e^{{i A_s}/\hbar}$ and $\psi_d= \sqrt{\rho_d} \times e^{{i A_d}/\hbar}$. The velocity fields of the ($s$) and ($d$) quantum fluids are given by $V_s=(\hbar/m) \nabla {A_s/\hbar}$ and $V_d=(\hbar/m) \nabla{ A_d/\hbar}$.  

Following Eq's. (\ref{eq.cont}) and (\ref{eq.Euler}), we can write the imaginary part of Eq.~(\ref{eq.psispsidschrodinger}) as a continuity equation and the derivative of its real part as a Euler equation with its two quantum potentials ($Q_s$ and $Q_d$). 
\beq
\frac{\d V_s}{\d t} + V_s. \nabla V_s =- \frac{\nabla \phi}{m} -\frac{\nabla Q_s}{m}-\l(\frac{\d V_d}{\d t} + V_d. \nabla V_d + \frac{\nabla Q_d}{m}\r)
\label{eq.54}
\eeq
\beq
\frac{\d \rho_s }{ \d t} + \text{div}(\rho_s V_s)=-\frac{\d \rho_d }{ \d t} - \text{div}(\rho_d V_d).
\label{eq.55}
\eeq
However, the ($d$) component of the quantum fluid, not involved in superconductivity, remains essentially static, so that $V_d=0$ and $\d \rho_d /{ \d t}=0$. Therefore we write a new system of fluid equations for the conducting quantum fluid ($s$):
\beq
\frac{\d V_s}{\d t} + V_s. \nabla V_s =- \frac{\nabla \phi}{m} -\frac{\nabla Q_s}{m}- \frac{\nabla Q_d}{m},
\label{eq.vsEuler}
\eeq
\beq
\frac{\d \rho_s }{ \d t} + \text{div}(\rho_s V_s)=0.
\label{eq.vscont}
\eeq
Re-integrating these equations under the form of a Schr\"odinger equation
\beq
\frac{\hbar^2}{2m} \Delta \psi_s + i \hbar \frac{ \d \psi_s}{\d t}- (\phi + Q_d)  \psi_s =0,
\label{eq.psisSchrodinger}
\eeq
we describe the motion of electrons ($s$), represented by their wave function $\psi_s$, in a potential well given by the exterior potential $\phi$, but also by an additional quantum potential $Q_d$, which is dependent on the local fluctuations of the density $\rho_d$ of static charges,
\beq
 Q_d= -\frac{\hbar^2}{2m} \frac{\Delta \sqrt{\rho_d}}{ \sqrt{\rho_d}}.
\label{eq.qd}
\eeq
The quantum potential well involves bound states in which two electrons can be trapped with zero total spin ($s_1+s_2=0$) and momentum ($p_1+p_2=0$). Under these conditions, $Q_d$ can be regarded as an interior potential with respect to $\psi_d$ but an exterior potential for $\psi_s$, justifying its inclusion alongside $\phi$ in the potential energy of the Schr\"odinger equation Eq.~(\ref{eq.psisSchrodinger}).  
\begin{figure}[h]
\begin{center}
\includegraphics[width=12cm]{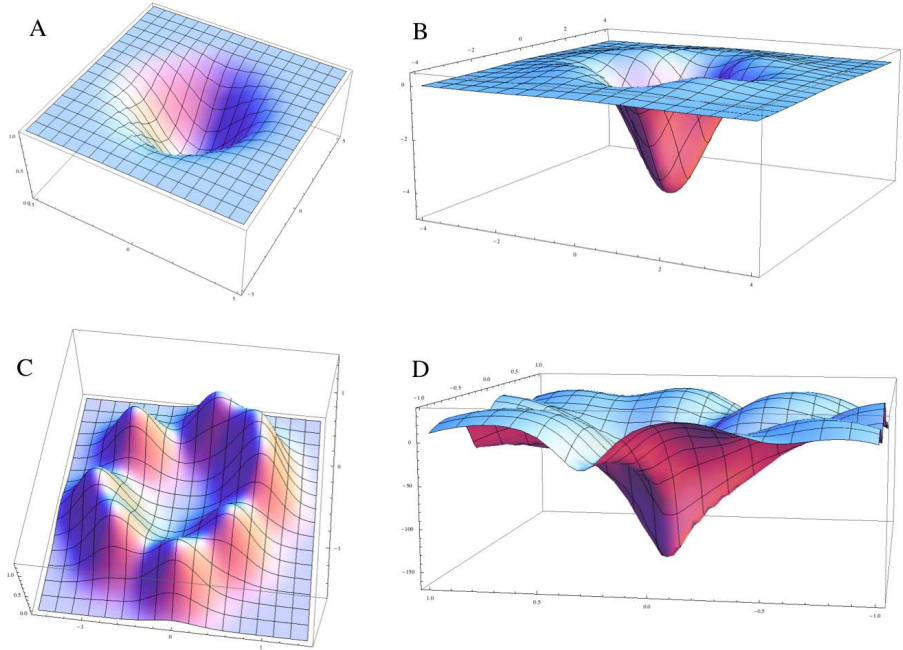} %
\caption{\small{({\bf A}) A simplified model of hole in the dopant distribution, simulated by the sum of four gaussians removed from the average dopant density. ({\bf B}) The $\approx 30\mathring{A}$ diameter quantum potential well computed from the idealized dopant density distribution of Fig. {\bf 3A}.  ({\bf C}) A more realistic model of dopant density distribution, in which a gaussian contribution to the density has been attributed to 8 dopant sites surrounding a void zone (due to ramdom fluctuations) corresponding to 4 missing dopant sites.  ({\bf D}) The quantum potential well computed from the dopant density distribution of Fig. {\bf 3C}. }} 
\label{fig3}
\end{center}
\end{figure}
The potential well (approximated by a anharmonic oscillator and known energy levels) can be computed from the dopant density distribution made from the sum of the gaussians.  It must be deep enough and have a correct shape to harbour two charges in an e-pair configuration ($p_1+p_2=0, s_1+s_2=0$). Simple models show that this ideally occurs for 4 dopants (8 charges).   As visualized in Figure~\ref{fig3}A, the 4 dopants create a  central void which equates to a $30\mathring{A}$ diameter potential well illustrated in Figure~\ref{fig3}B, of sufficient depth and of an appropriate shape to hold an e-pair in a bound state \cite{Nottale2014,Nottale2014b}.  This configuration of 4 dopant defects (oxygen atoms), gives a ratio $\rho_s/\rho_n = 2/(8 + 2) = 0.2$ at optimal doping, which coincides with experimentally  observed proportions of mobile dopants ($\approx 20\%$) over a range of different materials \cite{Tanner1998}.  

At optimal doping (Figure~\ref{fig2}), $p\sim 0.155 = 1/6.5$.  The 8 to 10 charges present in the potential well correspond to a surface $(8 - 10)\times 6.5 = (52 - 65) = (7.2 - 8.1)^2$.  As the lattice parameter $a$ of the $CuO_2$ planes = $3.8{\mathring{A}}$, this explains the mean size of the LDOS wells of $\approx {30}{\mathring{A}}$ observed experimentally \cite{McElroy2005}.  Taking this as an average scenario, it supports the hypothesis that the $Q_d$ coupling mechanism is linked to the SC phase at optimal doping, which connects with the peak of the parabolic curve in Figure~\ref{fig2}. 

We recognize that Figure~\ref{fig3}A represents an idealized case.  In reality there are a range of possible well configurations.  A detailed study of the impact of dopant levels on the formation of potential wells and e-pair coupling falls outside the scope of the present paper and will form the subject of a specific study in future work.  However, for illustrative purposes, an example of a more realistic configuration is given in Figure~\ref{fig3}C, along with its computed quantum potential Figure~\ref{fig3}D.  

We find that the quantum potential wells appear in the zones which are (because of random fluctuations) devoid of dopant defects, and therefore we expect the e-pairs created as bound states in these wells to lie preferentially in these zones. This is in accordance with the McElroy {\it{et al}} result \cite{McElroy2005} according to which bright regions with sharp coherence peaks, usually associated with strong superconductivity, occur between the dopant defect clusters.

We note that in under or over doping the number of dopant holes in a configuration suitable to hold e-pairs will be changed, leading to decline in e-pair DOS, correlation energy, $T^*$ and conductivity.   

We also note that the e-pairs trapped in the potential well are localized and that an additional mechanism is required to facilitate tunnelling between potential wells, which can explain macroscopic quantum coherence (and subsequent conduction), which we consider in sections 3.2.4 and 3.2.5.

\subsection*{{\it{3.2.3.2. The PG phase}}}
As stated previously, at optimal doping, dopants are arranged in a scale free (fractal) network \cite{Fratini2010}, inducing fractal AF spin wave fluctuations or `fractal magnons' ($fr_{mg}$) \cite{Prester1999,Prester2001}.  We propose that $fr_{mg}$ offer two key (separate) contributions to the underlying mechanism of HTSC.  

As a first step we consider their role in e-pair coupling, supporting the view proposed by Prester \cite{Prester1999,Prester2001}, that ${fr_{mg}}$ contribute to e-pair coupling.  However, whilst we do not preclude a role for fracton-coupled e-pairs in the SC phase, for reasons considered below, we preferentially identify this mechanism with the higher energy, localized PG phase ($\psi_{pg}$), prior to its merger with the SC phase.   

As discussed in detail by Savage and Affleck \cite{Savage1999}, a magnon corresponds to a soliton-antisoliton boundstate ($\psi_{mg}$), held together by a linear potential.  As in the case of fractal phonons considered by B\"uttner and Blumen \cite{Buttner1987}, the complex velocity field created by a fluid of fractal spin wave fluctuations ($fr_{mg}$), from which the wave function $\psi_{{fr}_{mg}}$ and its associated quantum potential Eq.~(\ref{eq.qmg}) originates, should actively enhance e-pair coupling, when compared to a linear potential\footnote{A full description of this mechanism falls outside the scope of the present paper and will be described in future work}. 
\beq
Q_{{fr}_{mg}}= -\frac{\hbar^2}{2m} \frac{\Delta \sqrt{\rho_{fr_{mg}}}}{ \sqrt{\rho_{fr_{mg}}}}.
\label{eq.qmg}
\eeq
As with Eq.~(\ref{eq.qd}), Eq.~(\ref{eq.qmg}) is analogous with Eq.~(\ref{eq.Q}).  However both equations fall within the standard description of quantum mechanics, based on $\hbar$, rather than the more generalised form of Eq.~(\ref{eq.Q}).  We highlight this point to differentiate between standard (microscopic) quantum potentials responsible for e-pair coupling and the next step (3.2.4 and 3.2.5) in which we consider the origin of a macroscopic quantum potential required to support macroscopic quantum coherence. 

To assist in setting the background for the next stage we use the approach followed in Eq's (\ref{eq.psispsidschrodinger} - \ref{eq.psisSchrodinger}) to higlight the role of the fracton derived quantum potential $Q_{{fr}_{mg}}$ in the PG phase.  As a first step we separate the PG wave function $\psi_{pg}$ from the fracton wavefunction $\psi_{{fr}_{mg}}$ involved in its formation via e-pair coupling, i.e.
\beq
\frac{\hbar^2}{2m} \Delta \psi_{pg} + i \hbar \frac{ \d \psi_{pg}}{\d t}- \phi\;  \psi_{pg} =-\frac{\hbar^2}{2m} \Delta \psi_{{fr}_{mg}} - i \hbar \frac{ \d \psi_{{fr}_{mg}}}{\d t}+\phi \; \psi_{{fr}_{mg}},
\label{eq.psipgpsimgschrodinger}
\eeq
Deconstructing Eq.~(\ref{eq.psipgpsimgschrodinger}) in terms of a Euler and continuity equation reveals the two quantum potentials associated with $\psi_{{fr}_{mg}}$ and $\psi_{pg}$.
\beq
\frac{\d V_{pg}}{\d t} + V_{pg}. \nabla V_{pg} =- \frac{\nabla \phi}{m} -\frac{\nabla Q_{pg}}{m}-\l(\frac{\d V_{{fr}_{mg}}}{\d t} + V_{{fr}_{mg}}. \nabla V_{{fr}_{mg}} + \frac{\nabla Q_{{fr}_{mg}}}{m}\r)
\label{eq.Eulerpgmg}
\eeq
\beq
\frac{\d \rho_{pg} }{ \d t} + \text{div}(\rho_{pg} V_{pg})=-\frac{\d \rho_{{fr}_{mg}} }{ \d t} - \text{div}(\rho_{{fr}_{mg}} V_{{fr}_{mg}}).
\label{eq.contpgmg}
\eeq
The quantum potential,  $Q_{{fr}_{mg}}$ associated with fractal spin waves $\psi_{{fr}_{mg}}$ contributes directly to e-pair coupling, forming an integral part of the localized PG phase ($\psi_{pg}$ and $Q_{{fr}_{mg}}$).  Equations (\ref{eq.Eulerpgmg}) and (\ref{eq.contpgmg}) can thus be simplified to 
\beq
\frac{\d V_{pg}}{\d t} + V_{pg}. \nabla V_{pg} =- \frac{\nabla \phi}{m} -\frac{\nabla Q_{pg}}{m},
\label{eq.Eulerpg}
\eeq
\beq
\frac{\d \rho_{pg} }{ \d t} + \text{div}(\rho_{pg} V_{pg})=0,
\label{eq.65}
\eeq
leading to a simplified Schr\"odinger equation in which $\psi_{{fr}_{mg}}$ and $Q_{{fr}_{mg}}$ are incorporated within the PG wave function $\psi_{pg}$
\beq
\frac{\hbar^2}{2m} \Delta \psi_{pg} + i \hbar \frac{ \d \psi_{pg}}{\d t}-\phi\;  \psi_{pg}= 0,
\label{eq.pgschrodinger}
\eeq 
We are assisted in our view that the PG phase is fracton ($fr_{mg}$) coupled by H\"ufner {\it{et al}} \cite{Hufner2008}, who reported that thermal conductivity `unexpectedly' correlates with PG relations in Figure~\ref{fig2}.  This observation offers an interesting insight, as thermal conductivity is phonon driven in ordered structures, crossing over to fractal phonons ($fr_{ph}$) as disorder increases.  This might on one level indicate a potential role for phonons in e-pair coupling. However, as discussed in section 2.3, most evidence suggests that a key role is not the most plausible scenario.  We therefore suggest an alternative interpretation of the observed temperature/PG relationship \cite{Hufner2008}. 

When considering a fractal network, we predict a strong correlation between thermal conductivity and the PG (if it is $fr_{mg}$ coupled), since $fr_{mg}$ and $fr_{ph}$, are determined by the same geometry.  As the thermal conductivity/PG relationship is supported by observations on a robust data set \cite{Hufner2008}, it suggests that e-pair coupling in the PG is $fr_{mg}$ mediated.  Conversely a complete lack of correlation between thermal conductivity and the SC phase \cite{Hufner2008} suggests that fractons do not play a significant role in the SC phase, at least prior to its merger with the PG phase.

Fractal magnon ($fr_{mg}$) coupling identifies a clear relation between AF insulators and HTSC.  However, based on evidence already cited \cite{Lanzara2001,Gweon2004} we do not necessarily exclude a role for fractal phonons ($fr_{ph}$) originally proposed as an e-pair coupling mechanism \cite{Buttner1987}.  Indeed the possible role of both phonons and spin fluctuations has already been considered in earlier work \cite{Li2005}.  In addition we do not rule out alternative mechanisms in different types of materials, which may also be influenced by fractal networks in a similar manner to that proposed by Prester \cite{Prester1999,Prester2001}.   For example, Ma {\it{et al}} \cite{Ma2014}, discovered that orbital fluctuations in iron-based compounds induce strongly coupled polarizations that can enhance electron pairing. 

\subsection*{{\it{3.2.4. The origins of macroscopic quantum coherence}}}

In this section we outline a second, previously unconsidered role for fractal spin waves ($fr_{mg}$), not involved in e-pair coupling, which establishes a further, fundamental connection between the AF properties of the cuprates and macroscopic coherence.  

So far we have considered standard quantum mechanics in which two coupling mechnisms based on $\hbar$, separately account for the SC and PG phase.   However, the anticipated delocalization associated with both mechanisms mitigates against e-pair coherence at macrosopic scales required for superconductance.  

As noted previously, as $T$ increases, $\lambda_{deB}$ decreases, suggesting a potentially negative effect on coherence length.  Established theory \cite{Anderson1958} suggests that this reduction in coherence length would be exacerbated in a fractal network due to localization.  This is to some extent supported by the localized properties of the PG below a critical level of doping (Figure 2).  However, a contradiction exists in the macroscopically coherent SC phase, which we need to address.  We also need to account for e-pair LDOS (reported by Mc Kelroy {\it{et al}} \cite{McElroy2005}) trapped in quantum wells created by dopants $Q_d$, which are intrinsically linked with the fractal network, and therefore additionally constrained by disorder.  

The ability of e-pairs to tunnel out of a potential well or between geometrically constrained LDOS depends on the probability of finding a nearby potential fluctuation into which trapped particles can tunnel. This indicates that we have an unidentified source of potential energy supporting tunnelling in the SC phase, which leads to the high e-pair DOS and coherence lengths observed.  This is extended to the PG phase at close to optimal doping, where the SC phase appears to merge with the localized PG phase (Figure~\ref{fig1} and Figure~\ref{fig2}). 

We have outlined an approach whereby a quantum potential can emerge from a fractal space Eq.~(\ref{eq.Q}) which is not confined to standard quantum mechanics defined by $\hbar$ but can also take a macroscopic form based on $\Ds$.  

We now describe how the fractal geometry of the dopants $\psi_d$ in the $p$-type cuprates can facilitate the counter intuitive transition from disorder to macroscopic quantum coherence.   

We begin with a classical diffusion process described by a Fokker-Planck equation,
\beq
\frac{\d P}{\d t}+ {\rm div}(P v)=D \Delta P,
\label{eq.78}
\eeq
where $D$ is the diffusion coefficient\index{diffusion!coefficient}, $P$ the probability density distribution of the particles and $v[x(t),t]$ is their mean velocity.

When there is no global motion of the diffusing fluid or particles ($v=0$), the Fokker-Planck equation is reduced to the usual diffusion equation for the probability $P$,
\beq
\frac{\d P}{\d t}=D \Delta P.
\label{eq.79}
\eeq
Conversely, when the diffusion coefficient vanishes, the Fokker-Planck equation is reduced to the continuity equation,
\beq
\frac{\d P}{\d t}+ {\rm div}(P v)=0.
\label{eq.80}
\eeq
We now make a change of variable, where in the general case, $v$ and $D$ are a priori non vanishing, 
\beq
V=v-D \nabla \ln P.
\label{eq.81}
\eeq
We first prove that the new velocity field $V (x, y, z, t)$ is a solution of the standard continuity equation.  Taking the Fokker-Planck equation and replacing $V$ by its above expression, we find
\beq
\frac{\d P}{\d t} + {\rm div} (P V)=\{ D \Delta P-{\rm div}(P v)  \} +  {\rm div} (Pv)-D \,  {\rm div} (P\nabla \ln P).
\label{eq.82}
\eeq
Finally the various terms cancel each other and we obtain also the continuity equation for the velocity field $V$,
\beq
\frac{\d P}{\d t} + {\rm div} (P V)=0.
\label{eq.83}
\eeq
Therefore the diffusion term has been absorbed in the re-definition of the velocity field.

\subsection*{\it{Euler equation and diffusion potential}}

We now consider a fluid-like description of the diffusing motion and determine the form of the Euler equation for the velocity field $V$.
As a first step we consider the case of vanishing mean velocity.

We first calculate the total time derivative of the velocity field $V$ in the simplified case $v = 0$
\beq 
\frac{d V}{dt}= \l( \frac{\d}{\d t} + V. \nabla\r)V=- D\,\frac{\d}{\d t} \nabla \ln P + D^2 (\nabla \ln P. \nabla) \nabla \ln P.
\label{TDV}
\eeq
Since ${\d}\nabla \ln P /{\d t}=\nabla {\d} \ln P /{\d t}=\nabla (P^{-1}{\d} P /{\d t})$, we can make use of the diffusion equation so that we obtain
\beq
\l( \frac{\d}{\d t} + V. \nabla\r)V=- D^2 \l[   \nabla \l( \frac{\Delta P}{P}\r)-(\nabla \ln P. \nabla) \nabla \ln P  \r].
\label{eq.inter}
\eeq
In order to write this expression in a more compact form, we use the fundamental remarkable identity Eq.~(\ref{eq.RI}), where $\psi=R$
\beq
\frac{1}{\alpha} \; \nabla\left(\frac{\Delta  R^{\alpha}}{R ^{\alpha}}\right)=   \Delta (\nabla \ln R)+2\alpha (\nabla \ln R . \nabla )
(\nabla \ln R ).
\label{86}
\eeq
By writing this remarkable identity for $R=P$ and $\alpha=1$, we can replace $  \nabla ({\Delta P}/{P})$ by $ \Delta (\nabla \ln P)+2 (\nabla \ln P . \nabla )\nabla \ln P$, so that Eq.~(\ref{eq.inter}) becomes
\beq
\l( \frac{\d}{\d t} + V. \nabla\r)V=- D^2 \l\{   \Delta (\nabla \ln P)+(\nabla \ln P. \nabla) \nabla \ln P  \r\}.
\label{eq.87}
\eeq
The right-hand side of this equation comes again under the identity Eq.~(\ref{eq.inter}), but now for $\alpha=1/2$. Therefore we finally obtain the following form for the Euler equation of the velocity field $V$:
\beq
\l( \frac{\d}{\d t} + V. \nabla\r)V=- 2 D^2 \, \nabla \l( \frac{\Delta \sqrt{P}}{\sqrt{P}}\r).
\label{eq.88}
\eeq
This is a new fundamental result \cite{Nottale2008}. Its comparison with the quantum result in the free case Eq.~(\ref{eq.quantfree}) is striking
\begin{equation}
\l(\frac{\partial}{\partial t} + V.\nabla\r) V  = +2{\Ds}^2 \, \nabla \l(\frac{\Delta \sqrt{P}}{\sqrt{P}}\r).
\label{eq.quantfree}
\end{equation}
This result demonstrates a clear equivalence between a standard fluid subjected to a force field and a diffusion process with the force expressed in terms of the probability density at each point and instant.  

The `diffusion force' derives from an external potential
\beq
\phi_{\rm diff}=2D^2 {\Delta \sqrt{P}}/{\sqrt{P}}.
\label{eq.diffp}
\eeq			
which introduces a square root of probability in the description of a totally classical diffusion process.  The quantum force is the exact opposite, derived from the 'quantum potential', which is internally generated by the fractal geodesics 
\beq
Q/m=-2{\Ds}^2 {\Delta \sqrt{P}}/{\sqrt{P}}.
\label{eq.quantp}
\eeq	
We speculate that these two forms of potential energy exist and compete in all quantum systems described by a Schr\"odinger equation. When the Quantum potential Eq.~(\ref{eq.quantp}) exceeds the diffusive potential Eq.~(\ref{eq.diffp}) a quantized structure emerges, contrasting with the anticipated diffusion to a more thermodynamically favorable, chaotic system observed in classical systems.  This is valid whether we consider subatomic particles, atoms or macroscopic quantum phenomena such as HTSC.  It's a genuine theory of self organisation, the structuring force coming from the interior instead of being an exterior force $\phi$.  We note that a very similar view on decoherence (taking a different approach and using different terminology) has been expressed by Dolce and Perali {\cite{Dolce2014}} who consider gauge symmetry breaking in terms of the competition between quantum recurrence and thermal noise.

\subsection*{\it{The role of the fractal dopant network}}

The idea that a MQP could emerge from fractal network has been previously considered \cite{Nottale1993,Nottale2009,Nottale2011}.  This earlier work suggested that the development of technological applications of such a concept could be achieved in a medium that would be fractal on several decades of scale, with objects in such a medium having macroscopic quantum-fluid-type properties, not based on $\hbar$, but on a new macroscopic constant specific to the system.  The underlying principle is that the relation of a medium (“container”) to the objects it contains is similar to the relation of a space (or space-time) to its objects. The difference is in the universality of space-time geometry, while the geometry of a medium is a constraint on the motion of the objects it contains only in the limit where they remain inside it and linked to it. If the theory is correct, it's a demonstration of the link between quantum mechanics and the theory of general relativity, the macroscopic force, induced by a fractal network being directly comparable with gravitational force, originating in the geodesics of space-time as well as being directly analogous with the standard quantum potential. 

Against this background we outline how a fractal network of dopants can lead to a MQP that facilitates macroscopic quantum coherence.  We set the hypothesis that the macro-scale fractal distribution of static dopants ($\psi_{D}$), which are not involved in e-pair coupling collectively creates a fractal electromagnetic scaffold through its associated MQP ($Q_D$), i.e.
\beq
Q_{D}  = -2{\Ds}^2 \, \frac{\Delta \sqrt{\rho_{D}}}{\sqrt{\rho_{D}}}.
\label{eq.QD}
\eeq
This concept raises an interesting question.  If holes between a configuration of dopants can be visualized as a quantum potential well ($Q_d$) as illustrated in Figure~\ref{fig3}, how can we visualize the effects of a fractal distribution of dopants on a macroscale which leads to a MQP?  We answer this by taking an example of a real, apparently unconnected group of dopants observed by McElroy2005 {\it{et al}} \cite{McElroy2005}, shown in figure 4A.  
\begin{figure}[!ht]
\begin{center}
\includegraphics[width=14cm]{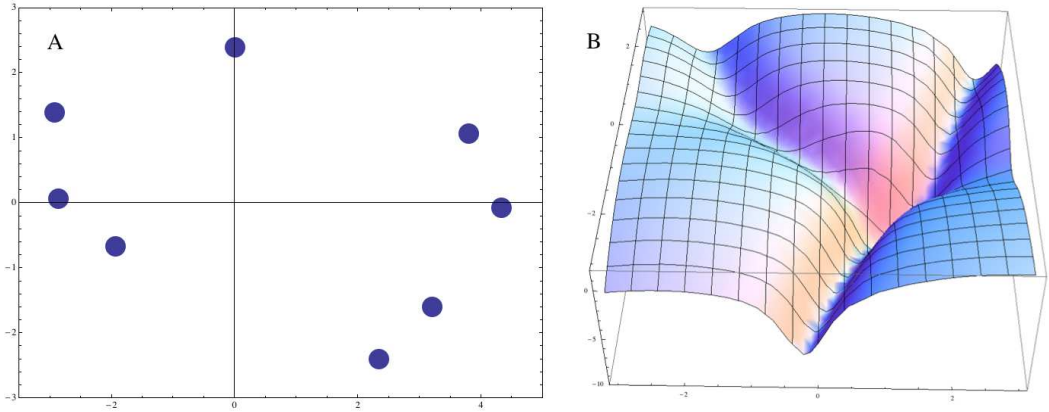} %
\caption{\small{({\bf A}) A sample of apparently unconnected dopants taken from McElroy{\it{et al}} \cite{McElroy2005}. ({\bf B}) A quantum potential $Q_d$ calculated from the dopant distribution of Fig. {\bf4 A} which is revealed as an undulating bifurcated channel. }} 
\label{fig4}
\end{center}
\end{figure}
If this dopant configuration is modelled following the approach described in section 3.2.3.1, the result is an undulating, bifurcated channel (Figure 4B) originating from a central (shallower) potential well rather than an isolated hole.  Starting from this image of bifurcation, it is not difficult to visualize how a fractal distribution of dopants would lead to a dynamic, fluctuating fractal network of channels connecting the deepest wells holding trapped e-pairs.  We conclude from this that $Q_D$ is very different from the simple sum of $Q_d$. 

The creation of a macroscopic web of this description offers an energetically favourable fractal network of paths channeling spin wave fluctuations $fr_{mg}$.  This system creates the environment, which through the mechanism described by (Eq's~\ref{TDV}-\ref{eq.quantp}), leads to coherent macroscopic spin wave fluctuations ($fr_{MG}$) .  

We denote these spin wave fluctuations  with a macroscopic wave function $\psi_{{fr}_{MG}}$ and an associated macroscopic quantum potential (MQP) $Q_{{fr}_{MG}}$.  This forms at a percolation threshold via the merger of quanta of fractal fluctuations $\psi_{{fr}_{mg}}$ and their associated quantum potentials ($Q_{{fr}_{mg}}$) Eq.~(\ref{eq.qmg}) across the cuprate lattice, i.e.
\beq
\l(\displaystyle\sum\limits_{n=1}^N {\psi_{{fr}_{mg_n}}}\,=\psi_{{fr}_{MG}}\r)\,\equiv\,\l(\displaystyle\sum\limits_{n=1}^N {Q_{{fr}_{mg_n}}}  \to Q_{{fr}_{MG}}  = -2{\Ds}^2 \, \frac{\Delta \sqrt{\rho_{fr_{MG}}}}{\sqrt{\rho_{fr_{MG}}}}\r).
\label{eq.QMG}
\eeq

We propose that $Q_{{fr}_{MG}}$ is the previously unidentified potential energy, facilitating escape of e-pairs from localized DOS, associated with the SC and PG phase, whilst the fractal network of channels created by $Q_D$ offers an energetically favourable route for e-pair tunnelling between LDOS.  

This mechanism would be expected to reveal itself as LDOS, interspersed by a web of coherent, superconducting channels.  A description which reflects images published by McElroy {\it{et al}} \cite{McElroy2005}, which can be considered as microscopic snap shots of much larger fractal networks described by Fratini {\it{et al}} \cite{Fratini2010}.  

In the process already outlined through Eq's~\ref{TDV}-\ref{eq.quantp}, we suggest that $Q_{{fr}_{MG}}$ and the fractal network created by $Q_D$ also supports the transition to macroscopic phase coherence of e-pairs.  To illustrate how this transition works in practice we draw a parallel with CRL, where a MQP induced by a fractal network (associated with CRL) plays the role of an optical cavity such as a Fabry-Perot resonator.  At some critical point beyond the percolation threshold in a fractal network, an inserted photon is predicted to spread when its wavelength is smaller than the mean free path of the material \cite{Anderson1958}.  As the number of photons (DOS) increase (i.e. an increase in gain), so will the probability of photon interaction. When energy levels reach a critical point, destructive interference effects (which we associate with observed quasiparticle interference in HTSC \cite{Damascelli2003,Alldredge2012}) will cancel out most frequencies, leaving those matching the resonant frequencies dictated by the geometry of the fractal network to form a complex velocity field and an associated MQP, leading to the establishment of a coherent standing wave.  Photons escaping from the random (fractal) network will be observed as CRL which we equate directly with coherent superconductivity.  

To date, no one has yet explicitly reported the presence of a MQP in HTSC.  However, as already indicated, evidence of $Q_D$ can be found in the presence of conducting channels between dopant clusters observed by McElroy {\it{et al}} \cite{McElroy2005}.  In addtion we speculate that $Q_{{fr}_{MG}}$ may be associated with a third harmonic energy scale signature observed by Alldredge {\it{et al}} \cite{Alldredge2012}.  It appears to be  particularly relevant as they reported a gradual increase in energy levels with increased doping before merging with the SC and PG phase at optimum doping. At this point, quasiparticle interference patterns associated with the energy scale disappeared. 

\subsection*{\centering{{\it{3.2.5. Relations between dopant levels and the emergence of macroscopic quantum coherence}}}}

In considering a theory to fully explain the role of the pseudogap and the fracton derived MQP in HTSC, we now consider relations between dopant levels and experimental observations reported in Figure~\ref{fig2}.  

At low levels of doping, repulsive forces between dopants will be minimal.  In an unconstrained system, they should be expected to self assemble in the most thermodynamically favorable state (a fractal network) during the annealing stage of material preparation \cite{Fratini2010}.  In the case of a base fractal granular structure \cite{Chadzynski2008}, this should have an additional impact on dopant self assembly, leading to increased levels of dopant disorder and it's $D_F$ when compared to a homogenous structure.  We anticipate that this effect should be enhanced as the $D_F$ of the granular structure increases.  We also expect local variation in structural $D_F$ around a specific mean value, with regions of highest disorder associated with higher energy fractal fluctuations ${\Ds}$ and the emergence of regional (still localized) $Q_{{fr}_{MG}}$ based quantum potentials, which we associate with the higher energy PG, below optimal doping.  In addition, the positive correlation between higher fracton fluctuations, which lead to higher energy ($Q_{{fr}_{mg}}$) coupling energies, should be reinforced by increasing temperature which we expect to act as a covariant in the temperature/PG relations observed in Figure~\ref{fig1} and Figure~\ref{fig2}.

When the fractal network created by increasing levels of dopants reaches a `percolation threshold' ($q_c$), we predict an infinite connected web of channels and fractal spin wave fluctuations.  At this point we expect the previously regionalised quantum potentials to have spread and merge into a larger MQP across the extent of the fractal network.  We identify this Quantum Critical Point with the transition from AF insulator to conduction observed in Figure~\ref{fig2}.  This interpretation is supported by differential conductance images using spectroscopic imaging scanning tunnelling microscopy \cite{Kohsaka2012} showing the formation of localized nano meter scale PG regions which begin to converge with increased doping, reaching a percolation threshold at a doping level of x=0.08, matching the data reported by Hufner {\it{et al}} \cite{Hufner2008} in Figure~\ref{fig2} for the onset of conduction.

When we consider the formation of potential wells (section 3.2.3.1) created by holes in spaces between dopants, the probability of their formation in homogenous materials is expected to be low at low levels of doping.  However, the probability of $Q_d$ formation is enhanced in areas of high $D_F$ where dopants may concentrate.  As doping levels increase, this probability increases further, leading to establishment of a SC phase on formation of the MQP at the percolation threshold ($q_c$).  As doping levels increase beyond this point we would expect to see continued increase in the e-pair DOS, the energy of the MQP and $T_c$ as seen in Figure~\ref{fig2}.  

At some point on the dopant level continuum beyond $q_c$ (depending on material, geometries and process conditions), equilibrium between dopant repulsion and the thermodynamic drive to fractal organization will be reached.  This is expected to lead to a MQP maxima which we associate with optimal doping, where tunnelling between local potential wells will peak.  At this point, e-pairs associated with the PG and SC phase will merge into a coherent condensate, maximizing DOS, correlation energy and coherence lengths, explaining the peak in $T_c$ observed (Figures 1 and 2).  

As doping increases beyond an optimum level, dopant repulsion will exceed the thermodynamic tendency to fractal organization.  At the upper limit this should lead to a homogenous doping network (supported by experimental observation \cite{Fratini2010}) and a transition from fractons to phonons and magnons.  During this transition, increased doping beyond optimal levels will lead to a decrease in the number of dopant induced potential wells, in a configuration appropriate to facilitate e-pair coupling in the SC phase.  Increasing dopant homogeneity will also lead to decline in the strength of fractal spin wave fluctuations, a reduction in $Q_{{fr}_{mg}}$ coupled e-pairs and the decline of the MQP, resulting in the eventual localization of any remaining $Q_d$ coupled e-pairs.  At this second quantum critical Point, the SC properties of the material will disapear (see Figure~\ref{fig1} and Figure~\ref{fig2}) as the material transforms into a Fermi liquid metal.

Depending on the point on the phase diagram (separate or merged phases) the Schr\"odinger equation will take different forms.  Between the onset of SC and optimal doping, $\psi_s$ transforms from a microscopic scale $\psi_s = \sqrt{\rho_s} \times e^{{i A_s}/\hbar}$ (where $A_s$ is a microscopic action), to a macroscopic scale, i.e.  
\beq
\displaystyle\sum\limits_{n=1}^N\psi_{s_n} \to  \psi_{S} = \sqrt{\rho_{S}}\times e^{i A_S/2\Ds},
\label{eq.68}
\eeq
where $A_S$ is a macroscopic action and m=1 (see section 3.2.2.3).

Up to this point the PG phase ($\psi_{pg} = \sqrt{\rho_{pg}} \times e^{{i A_{pg}}/\hbar}$) remains localized and the two phases  do not appear to interact.  However, at optimal doping, the PG phase also becomes macroscopic,
\beq
\displaystyle\sum\limits_{n=1}^N\psi_{pg_n} \to  \psi_{PG} = \sqrt{\rho_{PG}}\times e^{i A_{PG}/2\Ds}.
\label{eq.69}
\eeq
Beyond this point $\psi_{S}$ and $\psi_{PG}$ merge into a single macroscopic condensate, which we represent by a single wave function $\psi_C$, i.e.
\beq
\psi_C = \sqrt{\rho_{C}}\times e^{i A_C/2\Ds}.
\label{eq.71}
\eeq
As in Eq's (\ref{eq.psipgpsimgschrodinger}-\ref{eq.contpgmg}), to clarify the roles of $Q_D$ and $Q_{{fr}_{MG}}$ in $\psi_C$ we first separate their respective contributions, i.e.
\begin{multline}
{\Ds}^2\Delta \psi_{C} + i {\Ds} \frac{ \d \psi_{C}}{\d t}- \phi\;  \psi_{C} =-{\Ds}^2\Delta \psi_{{fr}_{MG}}  - i {\Ds} \frac{ \d \psi_{{fr}_{MG}}}{\d t}+\phi \; \psi_{{fr}_{MG}}\\-{\Ds}^2\Delta \psi_{D}- i {\Ds} \frac{ \d \psi_{D}}{\d t}+\phi \; \psi_{D},
\label{eq.CMGschrodinger1}
\end{multline}
which when written in terms of a Euler and continuity equation reveals the three quantum potentials ($Q_C$, $Q_D$ and $Q_{{fr}_{MG}}$)

\begin{multline}
\frac{\d V_{C}}{\d t} + V_{C}. \nabla V_{C} =- \frac{\nabla \phi}{m} -\frac{\nabla Q_{C}}{m}-
\l(\frac{\d V_{{fr}_{MG}}}{\d t} + V_{{fr}_{MG}}. \nabla V_{{fr}_{MG}} + \frac{\nabla Q_{{fr}_{MG}}}{m}\r)\\ 
-\l(\frac{\d V_{D}}{\d t} + V_D. \nabla V_D + \frac{\nabla Q_D}{m}\r)
\label{eq.CMGEuler1}
\end{multline}
\beq
\frac{\d \rho_{C} }{ \d t} + \text{div}(\rho_{C} V_{C})=-\frac{\d \rho_{{fr}_{MG}} }{ \d t} - \text{div}(\rho_{{fr}_{MG}} V_{{fr}_{MG}})-\frac{\d \rho_D }{ \d t} - \text{div}(\rho_{D} V_D).
\label{eq.CMGcont1}
\eeq

$\psi_D$ plays a key, but indirect role as an electromagnetic scaffold, created from the multiscale network of short range charge fluctuations, inducing fractal spin-waves ($fr_{mg}$) that merge to form macroscopic spin wave fluctuations ($fr_{MG}$) leading to $\psi_{{fr}_{MG}}$ and $Q_{{fr}_{MG}}$.   Without  $\psi_D$, $\psi_{{fr}_{MG}}$ and $Q_{{fr}_{MG}}$ would not exist. Based on this, we conclude that $Q_D$ fits the desription of an exterior potential.  According to this interpretation, in analogy with Eq's (\ref{eq.vsEuler}-\ref{eq.vscont}), $\psi_D$ remains static, i.e. $V_D=0$ and $\d \rho_D /{ \d t}=0$.  

We recall that a proportion of micro scale fractal spin wave fluctuations (${{fr}_{mg}}$) do not contribute to $\psi_{{fr}_{MG}}$ and its quantum potential $Q_{{fr}_{MG}}$, as they have already been absorbed into $\psi_{pg}$ through their contribution to e-pair coupling.  

We therefore regard $\psi_{{fr}_{MG}}$ as a macroscopic standing wave (framed by $Q_D$), with an associated macroscopic quantum potential $Q_{{fr}_{MG}}$.   From here we have two possible interpretations of the role of $Q_{{fr}_{MG}}$ in $\psi_C$.  The first, that it is seen as analogous to $Q_D$.  In this case it would act as an exterior potential facilitating tunnelling between localized e-pairs (both SC and PG phase) to create a coherent macroscopic phase, whilst not directly contributing to $\psi_C$.  According to this interpretation, Equations (\ref{eq.CMGEuler1}) and (\ref{eq.CMGcont1}) can be simplified to 
\beq
\frac{\d V_{C}}{\d t} + V_{C}. \nabla V_{C} =- \frac{\nabla \phi}{m} -\frac{\nabla Q_D}{m}-\frac{\nabla Q_{{fr}_{MG}}}{m},
\label{eq.CMGEuler}
\eeq
\beq
\frac{\d \rho_C }{ \d t} + \text{div}(\rho_C V_C)=0.
\label{eq.CMGcont}
\eeq

On re-integration under the form of a macroscopic Schr\"odinger equation, the exterior macroscopic quantum potentials $Q_D$ and $Q_{{fr}_{MG}}$ become explicit, whilst $Q_d$, present as an exterior potential in the microscopic equation Eq. (\ref{eq.psisSchrodinger}), disappears as it becomes internalised at the macroscopic scale.  

\beq
{\Ds}^2 \Delta \psi_C + i {\Ds} \frac{\partial\psi_C}{\partial t} - \l(\frac{\phi+Q_D+Q_{{fr}_{MG}}}{2}\r)\psi_C = 0.
\label{eq.SchrodingerCQ}
\eeq

The second interpretation is that $\psi_{{fr}_{MG}}$ and its associated quantum potential $Q_{{fr}_{MG}}$ plays a more dynamic, integral part of the conducting fluid $\psi_C$.  In this case ($\psi_{{fr}_{MG}}$ and $Q_{{fr}_{MG}}$) become absorbed into $\psi_C$ and Eq. (\ref{eq.SchrodingerCQ}) becomes 
\beq
{\Ds}^2 \Delta \psi_C + i {\Ds} \frac{\partial\psi_C}{\partial t} - \l(\frac{\phi+Q_D}{2}\r)\psi_C = 0.
\label{eq.SchrodingerC}
\eeq

At this moment in time it is not possible to state with certainty whether $Q_{{fr}_{MG}}$ should be regarded as an internal or external potential.  Further work is required to better understand this.  We also need a better understanding of the importance of charge order and currently undefined additional factors represented by $\phi$, such as magnetic fields, a non-linear term of the Ginzburg-Landau type and/or additional couplings.

Based on what has been described above, we suggest that Figure~\ref{fig1}(b) is a reflection of the coupling mechanisms described, whilst Figure~\ref{fig1}(c), an example of which is published by Norman \cite{Norman2010}, better reflects the e-pair phase relations at optimal doping described by Eq's (76-77). We suggest that the discrepancies between Figure~\ref{fig2}  and Figure~\ref{fig1}(b) and Figure~\ref{fig1}(c), may result from variation associated with the merger of multiple data sets in Figure~\ref{fig2}.

{\textbf{Additional observations}}
We have attributed fractal spin waves as a key contributor to macroscopic quantum coherence.  However, we have not yet considered the role of other forms of electronic order such as charge and spin density waves, which until recently were not considered as playing an important role in HTSC \cite{Fradkin2012}.  However, there is an increasing body of experimental evidence to suggest that they should be considered as part of the picture \cite{Fradkin2012}.  Making sense of this is not a simple task due to the wide range of materials of different composition, exhibiting different levels of charge order.  

A detailed study of how charge order fits into our proposed theory falls outside the scope of the present paper.  However, there are a number of signposts to assist in formulating a tentative hypothesis.  Fradkin and Kivelson \cite{Fradkin2012} posed the question `is local Charge Density Wave (CDW) ordering the cause of the PG or a derivative phenomenon that can arise from PG correlations?'  

It seems reasonable to propose that dopant order, which can induce fractal spin wave fluctuations could also influence charge order with the expectation of synergystic effects on the MQP ($Q_{{fr}_{MG}}$). This idea is supported by Comin {\it{et al}} \cite{Comin2014} who have reported a relationship between dopant levels and charge order wave vectors.  Their work suggests an intimate relationship between charge order and the PG, which we associate with fractal spin wave fluctuations and the resulting MQP.  The hypothesis is supported by Fujita {\it{et al}} \cite{Fujita2014} who reported electronic symmetry breaking in the underdoped cuprates, which disapears in the overdoped region $(x\approx0.19)$.  In Figure 1A of their paper, relations between symmetry breaking and dopant level, precisely match the dopant/PG relations shown in Figure~\ref{fig1}b.  This indicates that electron wave fluctuations are being influenced in a similar manner to spin wave fluctuations. This links at least in part to observations by Poccia {\it{et al}} \cite{Poccia2012} on dopant induced strains on the embedding medium, leading to a fractal distribution of local lattice distortions, which they equate with static CDW's.  We therefore conclude that the onset and disapearance of symmetry breaking reported by by Fujita {\it{et al}} \cite{Fujita2014} is linked to quantum critical points, which we have identified with the onset and subsequent breakdown of the fractal dopant network and its impact on $Q_D$ and $Q_{{fr}_{MG}}$.

On an additional note, Fujita {\it{et al}} \cite{Fujita2014} have made a connection between weakening of electronic symmetry breaking (at over doping) with closure of antinodal spacing between the Fermi arcs.  This relates directly to observations by Comin et al \cite{Comin2014} who have explored the link between charge order and Fermi arc instability.  Based on these observations, we speculate that Fermi arc instability may be linked to e-pair tunnelling across the cuprate lattice via the antinodes (offering the path of least resistance), which has a direct relationship with dopant order, charge order and the MQP's ($Q_D$ and $Q_{{fr}_{MG}}$).

To confuse matters it has been reported that charge density waves potentially compete with SC \cite{Chang2012}.  This is based on the observation that below $T_c$, a strong magnetic field leads to a stronger CDW amplitude combined with reduced SC.  The theory we propose offers new insight on this observation.  A strong magnetic field will destroy magnetic spin wave fluctuations (and therefore $Q_{{fr}_{MG}}$) leading to a breakdown of SC.  At the same time, any synergistic effects on CDW order (which we have linked to symmetry breaking) influenced by $Q_{{fr}_{MG}}$ would also be destroyed, with CDW order reorienting within the external magnetic field.  This interpretation suggests that charge order is not fundamentally in competition with SC under normal conditions (in the absence of a strong external magnetic field).

To summarise our tentative position on charge order.  There is strong evidence to suggest that the fractal arrangement of dopants which induces fractal spin waves also induces a synergistic arrangement of charge order.  Comin {\it{et al}} \cite{Comin2014} referenced several papers indicating the presence of both short and long range charge correlations in different materials.  A picture is beginning to emerge of charge order being influenced by $Q_D$, which may act synergistically with $Q_{{fr}_{MG}}$ to create a multicomponent complex electromagnetic velocity field (and an associated MQP), which may lead to a positive correlation between long range charge order and $T_c$. 

As noted at the outset, the theory outlined in the paper is focussed on the hole doped ($p$-type) cuprates which have been studied most extensively.  Further work is required to determine which components of the current approach are relevant for other HTSC materials exhibiting different characteristics.  However, there do appear to be some common threads.  In one example, the phase diagram linked to electron doped ($n$-type) cuprates differs in a number of respects from the $p$-type cuprates \cite{Silva2014,Yeh2002}.  One notable difference is the shorter dopant range over which the SC dome exists \cite{Silva2014,Yeh2002}.  Two immediate ideas that may account for this come to mind. The first relates to electron doping in the $d$-orbital of $Cu$ in $n$-type, versus hole doping in the oxygen $p$-orbital of $p$-type cuprates.  This gives rise to spinless $Cu^+$-ions that dilute the background AF $Cu^{2+}-Cu^{2+}$ coupling, resulting in weaker spin fluctuations compared with the $p$-type cuprates  \cite{Yeh2002}.  It is reasonable to expect that formation of $Q_{{fr}_{MG}}$ would be more constrained under these circumstances.  The second idea relates to the $Q_d$ mechanism for e-pair coupling.  It is theoretically possible for electrons doped into the AF insulator to form potential wells as described in Section 3.2.3.1.  However, geometries will differ and the probability of this occuring will be significantly reduced, particularly in the underdoped region.  More experimental data and modelling work is required to confirm this hypothesis.  However, if such a mechanism exists, it would certainly translate to a smaller dopant range supporting SC, which is reflected in the smaller SC dome found in the $n$-type cuprates.

In a second example, it was until recently thought that charge order, (commonly observed in the $p$-type cuprates) did not exist in the $n$-type cuprates.  However,  recent results have challenged this position \cite{Silva2014}, indicating that charge order in $Nd_{2-x}Ce_xCuO_4$ exhibits similar periodicity and orientation to the p-type cuprates.  However, the onset of charge order is higher than the PG temperature in the region where AF fluctuations are first detected \cite{Silva2014}.  da Silva Neto {\it{et al}} \cite{Silva2014} conclude from this that a much stronger link exists between charge order and spin wave fluctuations.  The concept is not incompatible with our theory of dopant induced fluctuations (associated with spin wave and charge order), where a different dynamic is to be expected with different dopants.

The two examples given above suggest that some components of our proposed theory may share commonality with HTSC materials beyond the $p$-type cuprates.  However, given the extensive range of experimental data relating to different HTSC materials it is not practical to comment on all observations and theoretical approaches in the current paper.  As previously stated, we do not rule out alternative e-pair coupling mechanisms which may  contribute to SC, particularly when supported by data conflicting with that used to develop our current theoretical approach.  At the same time, we speculate that a common trait in hetergenous HTSC materials is the establishment of disorder induced resonances, leading to MQP facilitated e-pair tunnelling and coherence.  

We hope that the current theory will be seen as a platform upon which further ideas can be developed to shed light on variations in experimental data available on a growing range of HTSC materials. 

\subsection*{4.  Conclusions}
We have proposed a new, evidence based theoretical approach to macroscopic quantum coherence and superconductivity in the most studied family of HTSC materials; the $p$-type cuprates.  The theory addresses the key challenges outlined in section 3.1.

As a first step we consider mechanisms to explain e-pair coupling in the superconducting and pseudogap phases and their inter relations.  

Electron pair coupling in the superconducting phase is facilitated by dopant induced potential wells in a hypothesis confirmed by independent expermental data \cite{McElroy2005}.  Our interpretation is aligned with experimentally observed optimal doping levels and the peak in $T_c$.  By contrast, electrons in the pseudogap are coupled by fractal spin waves (fractons) induced by the fractal arrangement of dopants.  

On another level, the theory offers new insights into the emergence of a macroscopic quantum potential generated by a fractal distribution of dopants ($Q_D$).  This, in turn, leads to the emergence of coherent, macroscopic spin waves and a second associated quantum potential $Q_{{fr}_{MG}}$, possibly supported by charge order.  These quantum potentials play two key roles.  The first involves the transition of an expected diffusive process (normally associated with Anderson localization) in fractal networks, into e-pair coherence.  The second involves the facilitation of tunnelling between localized e-pairs.  These combined effects lead to the merger of the super conducting and pseudo gap phases into a single coherent condensate at optimal doping.  The underlying theory relating to the diffusion to quantum transition is supported by Coherent Random Lasing, which can be explained using an analogous approach. 

The paper highlights a key difference between macroscopic quantum coherence in conventional SC and HTSC.  The origins of conventional SC is driven by e-pair coupling with large coherence length related to standard quantum mechanics  by $\lambda_{deB}=\hbar/mv$, where $\lambda_{deB}$ grows to macroscopic levels at very low $T$.  We contrast this with HTSC, which is facilitated by different coupling mechanisms combined with a macroscopic quantum potential, derived from the structure of the media which means $\hbar$ is replaced with $2m{\Ds}$, i.e.  $\lambda_{deB}=2{\Ds}/{v}$.

\subsection*{5.  Future work}
Our theoretical approach offers new insights into how to improve $T_c$ in heterogeneous HTSC.  If we increase the $D_F$ of the granular base material beyond existing levels it should theoretically lead to an increase in the $D_F$ of doping geometries (for a specific set of process conditions \cite{Poccia2012,Poccia2011b}) and an associated increase in fractal fluctuations $\Ds$ and MQP strength.  This is predicted to enhance tunnelling between the SC and PG phase, reducing localization of the PG, leading to increasing e-pair correlation energies and $T_c$.

This could be tested by manufacture of different materials with granular micro structures of higher $D_F$.  Such a program should also consider the impact of different quasiparticle/excitation based fractons with different frequencies and energies in different materials, which are expected to influence e-pair coupling energies.  As part of this work, observed trends in different materials such as $T_c [Hg] > T_c [T1] > T_c [Bi]$ reported by Pickett \cite{Pickett2006}, should be investigated.  This work should also consider extending the scale of fractal networks beyond the $400\mu m$ reported by Fratini et al {\cite{Fratini2010}}.  Studies by Poccia {\it{et al}} \cite{Poccia2011} indicate that extending the multiscale nature of dopant ordering leads to an increase in $T_c$.  We need to determine how far we can take this with new material structures.  This work should confirm or refute the theory that $T_c$ is related to the scale of the fractal networks.

Enhanced MQP effects are expected in symmetric $3D$ fractal structures (with increased correlation energies across the network) compared to $2D$ ordering of dopants.  This approach is already addressed to some extent by the multilayer structures found in HTSC materials \cite{Poccia2011} which creates a $3D$ connective network via interlayer tunnelling. However, interlayer discontinuities may account for previously unexplained phenomena \cite{Pickett2006} where $T_c$, increases with up to 3 $CuO_2$ layers, but then decreases with the addition of further layers \cite{Pickett2006}.  We speculate that the effects of multilayer discontinuity could interfere with MQP formation.  With new insights from this current work, we propose a fresh look at optimising $3D$ symmetries.  This approach requires an investigation into the possibility of creating continuous 3D fractal structures on a number of scales, annealing a wide particle size distribution along lines conceptually developed by Nottale \cite{Nottale2011},  and specifically proposed by Barthelemy et al \cite{Barthelemy2008}.  Demonstrating increased coupling energies and/or a significant reduction in localization of the PG would validate the proposed theory and identify opportunities for further development.  

When considering CRL, if the $D_F$ of a network is increased, it will increase the diffusive potential of the system.  This will require a greater photon DOS (higher levels of gain) before coherence is reached.  However, at this critical point, the correlation energy of the system (due to increased frequency and DOS) is predicted to be higher.  The higher frequencies could  easily be detected in future experimental work to confirm our hypothesis. 

Experimental work needs to be combined with a quantitative approach to assist in understanding the relative roles of the interacting components more specifically  and to identify critical areas of focus for on-going improvements.

If our theory is correct it extends established theory \cite{Anderson1958}, offering new insights to address localization constraints in disordered media.  Opportunities for future development extend well beyond the two concepts highlighted in this paper (CRL and HTSC),  with important implications for resolving decoherence challenges in quantum computing and understanding macroscopic phenomena observed in biological systems which have been previously discussed \cite{Poccia2011,Auffray2008,Nottale2008,Nottale2011}.  
\subsection*{Acknowledgments}
The authors would like to acknowledge the role of the COST office, and in particular COST Action FP1105 in facilitating the collaboration that led to this work.  We would also like to thank H\"ufner {\it{et al}} \cite{Hufner2008} for granting us permission to use figures 1 and 2 extracted from their paper.  Finally we would like to thank Andrea Damascelli for his constructive comments in reviewing a first draft of the paper.



\end{document}